\documentclass[12pt]{article}
\usepackage{graphicx}
\usepackage{cite}
\usepackage{booktabs}
\usepackage{array}
\usepackage{amsmath}
\usepackage{url}

\graphicspath{{./}{Figs/}}

\begin{document}

\title{Ambiguities, Built-in Biases and Flaws \\ in \\ Big Data Insight Extraction}

\author{Serge Galam\\ CEVIPOF - Centre for Political Research,\\
SciencesPo and CNRS, \\ 1, Place Saint-Thomas d'Aquin, 75007 Paris, France \\ June 25, 2025}
\date{serge.galam@sciencespo.fr}

\maketitle

\begin{abstract}

I address the challenge of extracting reliable insights from large datasets using a simplified model that illustrates how hierarchical classification can distort outcomes. The model consists of discrete pixels labeled red, blue, or white. Red and blue indicate distinct properties, and white represents unclassified or ambiguous data. A macro-color is assigned only if one color holds a strict majority among the pixels. Otherwise, the aggregate is labeled white, reflecting uncertainty. This setup mimics a percolation threshold at fifty percent. Assuming direct access of the various proportions of colors is infeasible from the data, I implement a hierarchical coarse-graining procedure. Elements (first pixels, then aggregates) are recursively grouped and reclassified via local majority rules, producing ultimately a single super-aggregate whose color represents the inferred macro-property of the collection of pixels as a whole.  Analytical results, supported by simulations, show that the process introduces additional white aggregates beyond white pixels, which could be initially present. These arise from groups lacking a clear majority, requiring arbitrary symmetry-breaking decisions to attribute a color to them. While each local resolution may appear minor and  inconsequential, their repetitions introduce a growing systematic bias. Even with complete data, unavoidable asymmetries in local rules are shown to skew outcomes. This study highlights a critical limitation of recursive data reduction. Insight extraction is shaped not only by data quality, but by how local ambiguity is handled. That results in built-in biases. Thus, the related flaws are  not due to the data, but due to structural choices made during local aggregations. Though based on a simple model, the findings expose the high likelihood of inherent flaws in widely used hierarchical classification techniques.

\end{abstract}

\section{Introduction}

Collecting and treating huge mass of data became an essential part of almost any activity, covering a rather large spectrum of fields. However, the related extraction of accurate and meaningful information from heterogenous and diverse datasets may turn delicate and sometime misleading  with risks of information loss, bias outcomes, and misinterpretation. That is a major challenge for data science. 

Yet, many common analytical techniques are available. However, while these techniques are designed to extract inferences under uncertainty, each one of them incorporates some specific trade-offs that can significantly affect the feasibility of the insights obtained.

For instance, Principal Component Analysis (PCA) is a widely used technique for dimensionality reduction, transforming high-dimensional data into a lower-dimensional space that captures the greatest variance \cite{jol}. However, PCA assumes linear relationships and can obscure localized or nonlinear structures. Similarly, t-distributed Stochastic Neighbor Embedding (t-SNE) is a nonlinear technique effective for visualizing aggregates, but it is highly sensitive to parameter choices and does not preserve global distances well \cite{maa}.

On another hand, aggregating algorithms such as k-means and DBSCAN (Density-Based Spatial Aggregating of Applications with Noise) aim to partition data into meaningful groups. While k-means minimize intra-aggregate variance assuming spherical aggregates of similar size, DBSCAN identifies aggregates based on density connectivity and can detect arbitrarily shaped aggregates \cite{est}. However, both methods require careful parameter tuning and can struggle with noisy or unbalanced data.

In addition, when multi-scales data are involved like in statistical physics and image analysis coarse-graining methods are very efficient in extracting insights  \cite{coa}. These methods can handle large-scale data, yet at each stage of aggregation, ambiguity resolution and information compression introduce the potential for distortion, especially when uncertainty is ignored or oversimplified \cite{gol}.

All these methodological issues become particularly critical in classification tasks, where simplified representations must still support accurate decision-making. Among a series of potential flaws, small distortions can cascade into systematic misclassifications. 

Indeed,  research in algorithmic fairness and explainable machine learning shows that oversights in preprocessing or model design can have epistemic and ethical consequences \cite{bar}. Therefore, extracting insights from a data set is not merely a technical process but a conceptual one-requiring scrutiny of the assumptions baked into every transformation, aggregation, and inference  \cite{dos, lip}. 

I demonstrated this point in an idealized case of identifying a would-be terrorist from a large set of data obtained by monitoring a suspicious person. By labelling each ground item as Terrorist Connected or Terrorist Free, a coarse-graining of all collected ground items is implemented to end up by eventually label the person under scrutiny as a would-be terrorist or a not would-be terrorist  \cite{would, tip}. The results showed the existence of systematic wrong labelling for some specific ranges of the item proportions. In particular, the flaw proves to be irremovable due to its anchor within the treatment of uncertain aggregates of items, which inevitable appear. 

In this paper, I extend the above illustration to a more general setting by addressing the challenge of extracting reliable macroscopic information from non-annotated microscopic data. To that end, I investigate a stylized model that determines the macro-color of a collection of individually colored pixels using a repeated coarse-graining process based on bottom-up hierarchical aggregation. Each pixel is assigned one of three colors, red, blue, or white. Red and blue serve as illustrative categories that can represent any form of underlying content or classification such as traits, behaviors, or detection statuses, depending on the application. In contrast, white denotes an unclassified or undecided state.

The process begins by forming groups of $r$ randomly selected pixels. The color of each aggregate is then determined by applying a local majority rule. The procedure is repeated across successive layers, with each new layer formed by grouping $r$ aggregates from the layer below and assigning their color using the same local majority rule. As the hierarchical structure develops, the system ultimately converges to a final super-aggregate encompassing all pixels. The resulting macro-color provides the final classification of the entire collection. When more than fifty percent of the pixels are red or blue, the corresponding macro-color is expected to be red or blue, respectively. Otherwise, it is considered white.

By fixing all aggregates to size $r=4$, I am able to solve the model analytically and complement the analysis with simulations. While the hierarchical scheme is intuitive and computationally efficient, the results reveal that it produces misleading outcomes. The repeated application of local majority rules introduces cumulative information loss, which becomes increasingly pronounced at higher levels of aggregation. In particular, early ambiguities caused by possible white pixels and tie groupings propagate upward and, through repeated filtering, always distort the final result for some specific range of the respective proportions of pixel colors. As a consequence, the system frequently converges to a macro-color that does not represent accurately the original proportions of red and blue pixels.

This distortion is not the product of randomness or local fluctuations but stems from structural limitations inherent in the coarse-graining process. The recursive majority rule acts as a nonlinear filter, suppressing white aggregates and amplifying early local biases. Even very few whites aggregates at iniital layers can disproportionately affect the outcome, leading to systematic misclassification. By analyzing how such distortions emerge and accumulate, the results exposes the subtle but significant flaws embedded in hierarchical aggregation methods commonly employed in data reduction and classification tasks.

Last but not least, it is worth noting that similar phenomena occur in opinion dynamics with the democratic spreading of minorities, which take advantage of doubts and prejudices to convince an initial majority to shift opinion  \cite{public}.The thwarting of rational choices is also active in financial markets \cite{inv}. These works subscribe to the active field of the modeling of opinion dynamics within the field of Sociophysics  \cite{r1, r2, r3, r4, r5, r6, r7, r8, r9, r10, r11, r12, r13, r14, r15}.

\section*{Outline of the Paper}

The remainder of the paper is organized as follows. In Section 2, I define the hierarchical coarse-graining procedure in precise terms and introduce the majority rule dynamics governing color assignment at each level.  Analytical tools are employed to characterize the evolution of color distributions through successive layers of aggregation, with special focus on the case $r = 4$.

An exact analytical solution is derived for the setting in which only blue  (B) and red (R) pixels are present.  The use of repeated local aggregations is shown to produce indecisive aggregates already at the first level of the hierarchy. In cases of ties, aggregate are labeled B with probability $k$ and R with probability $(1-k)$. The repeated appearance of local ambiguities with the related symmetry-breaking can drive the spread of the minority pixel color while climbing the hierarchy. 

The impact of including a third white color is studied in Section 3. Exact analytical solutions demonstrate how local ambiguities propagate through the formation of  white aggregates. I provide a detailed examination of the nonlinear filtering effect induced by repeated majority decisions and its consequences for the final macro-color classification.

In Section 4, I build out the full flow of colors leading to the macro-color for a series of different set of parameter values.

Section 5 reports the results of numerical simulations that complement the theoretical analysis. These simulations explore the probability of correct classification under varying initial conditions, such as different proportions of red, blue, and white pixels. They also quantify how small local fluctuations can lead to large-scale misclassifications.

A short discussion is provided in the Conclusion. In particular, I emphasize the major impact of local decision rules during coarse-graining introducing hidden biases. Understanding these structural flaws is critical for designing more robust and interpretable data aggregation procedures.

\section{Majority Rule and Ambiguity for two-color pixels}

In this section, I define the hierarchical coarse-graining procedure given a collection of individually colored pixels with respectively red (R) and blue (B) colors  in proportions $p_0$ and $(1-p_0)$.
 
All pixels are then distributed randomly in groups of 4 yielding $2^4=16$ different types of configuration. Among them, 2 are single colored and 8 are composed with 3 pixels of the same color and 6 have a tie with 2 R and 2 B. The 16 configurations reduce to 5 in terms of different compositions of R and B. 

Applying a majority rule to the first 10 configurations yields 5 full R and 5 full B. The last 6 tied configurations are undetermined from majority rule. They require a special handling. To cover all possible treatments of a tie configuration I introduce the parameter $k$, which provides the probability that 2 A 2 B $\rightarrow$ 4 A and $\rightarrow$ 4 B with probability $(1-k)$.

Each single-colored group is now turned to one aggregate of level one with the group color. The probability of having a level one R aggregate is,
\begin{equation}
p_{1}= p_0^4+4p_0^3(1-p_0)+6k p_0^2(1-p_0)^2 .
\label{p1} 
\end{equation}

Repeating the procedure to build aggregates of level $2, 3, \dots, n$ where last level n encompasses the full collection of pixels, lead to $p_0 \rightarrow p_1 \rightarrow p_2 \rightarrow ...\rightarrow  p_{n-1} \rightarrow  p_n$. For a collection of $N$ pixels, the number n of successive coarse-grained iterations is given by,
\begin {equation}
n=\left\lfloor \frac{ \ln N} {\ln 4} \right\rfloor ,
\label{n4} 
\end{equation}
where the cell part ensures an integer value for the number of hierarchical levels.

The Logarithmic dependence on N indicates that collecting larger numbers of items does not require a significant increase in the number of required iterations to treat the corresponding collection of pixels. For instance, going from $N=4096$ to $N=16384$, i.e., adding 12288 pixels requires only one additional coarse-graining with $n=7$ insated of $n=6$. To go up to $N=65536$ pixels requires $n=8$ and only $n=10$ to process the huge number $N=1.04858 \ 10^6$.

To determine the dynamics driven by iterating Eq.  (\ref{p1}), I solve the fixed point Equation $p_1=p_0$ to obtain the values  which are invariant under coarse-graining. Three fixed points are obtained with $p_{R}=1$, $p_{B}=0$ being attractors and
\begin{equation}
p_{c,k}=\frac{(1-6k)+\sqrt{13-36k+36k^2}} {6(1-2k)},
\label{k4} 
\end{equation}
a tipping point between them. With $p_{c,0}=\frac{1+\sqrt{13}} {6} \approx 0.77$,  $p_{c,1/2}=\frac{1}{2}$ and $p_{c,1} = \frac{5-\sqrt{13}} {6} \approx 0.23$, I obtain  $\frac{1}{2} \leq p_{c,k}\leq  0.77$ for $0\leq k \leq \frac{1}{2}$ and $0.23\  \leq p_{c,k}\leq \frac{1}{2}$for $\frac{1}{2}\leq  k \leq 1$.

At this stage it is worth stressing that getting the level-n aggregate does not guarantee to reach one of the two attractors with $p_n\neq p_R \ \& \ p_B$. In such a case the color labelling is probabilistic R with probability $p_n$ and B with probability $(1-p_n)$.

Starting from a proportion $p_0$ of R pixels, to ensure reaching of the two attractors requires a number,
\begin {equation}
m=\left\lfloor \frac{1}{\ln \lambda_k} \ln \frac{1}{\vert 1-\frac{p_0}{p_{c,k}}\vert} \right\rfloor +2 ,
\label{m} 
\end{equation}
of hierarchical levels \cite{geo} where $ \lambda_k=\frac{dp_1}{dp_0} \vert_{p_{c,k}}$ with $ \lambda_0= \lambda_1\approx 1.64$ and $\lambda_{\frac{1}{2}}=\frac{3}{2}$.

While Eq. (\ref{m}) is an approximation \cite{book}, it yields exact results for most cases. Most of the associated values are less than 10 as seen from  Figure (\ref{m}) for $k=0$. Only in the immediate vicinity of  the tipping point $p_{c,0}\approx 0.77$ does $m$ exhibit a cusp around 20. There, $N=4^{20}$ pixels are required to reach the attractor, which is often out of reach (Eq. (\ref{n4})). Therefore, given $p_0$ a minimum number $N=4^m$ pixels is required, otherwise the color labelling is probabilistic.

\begin{figure}
\includegraphics[width=.60\textwidth]{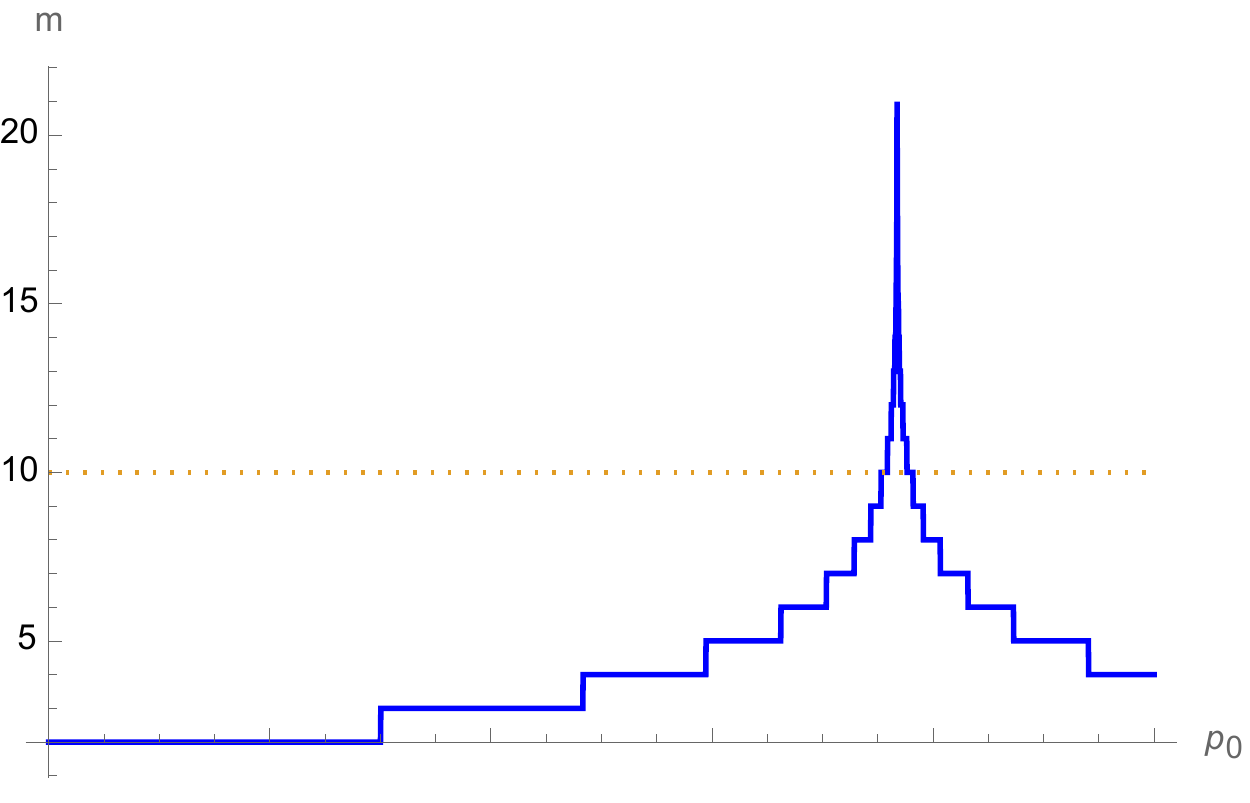}\
\caption{Number of iterations required to reach one of the two attractors and thus get a deterministic label of the giant item as a function of the proportion $p_0$ of ground items.}
\label{m}
\end{figure} 

Applying local majority rules to even size groups introduces naturally the challenge to deal with ambiguities in extracting information. These ambiguities appear naturally when a local  tie occurs randomly  with the same number of R and B.  In a aggregate of two R and to B pixels, there is no majority to identify the aggregate color, which in turn creates an ambiguity on how to color the related aggregate. A decision has to be done to select one of the two colors. 

Combining above results with the conditions $p_0>0.50 \Rightarrow$ macro-R (red) and $p_0<0.50 \Rightarrow$ macro-B (blue) proves a systematic error with a wrong macro-color in either one of the two following cases:

When $0\leq k \leq \frac{1}{2}$ the range $\frac{1}{2} \leq p_0\leq p_{c,k}$ yields a macro-B while the exact color is macro-R. 

Similarly, when $\frac{1}{2} \leq k \leq 1$ the range $p_{c,k} \leq p_0\leq \frac{1}{2}$ yields a macro-R while the exact color is macro-B. 

While these local wrong outcomes are meaningless per se, they are found to end up disrupting drastically the expected final outcome in term of the actual macro-color of the collection of pixels.

The different scenarios can be illustrated with the case of tagging a would-be terrorist a person under scrutiny \cite{}. Red pixels are then associated to data which are Terrorist Connected while blue pixels means Terrorism Free data. A macro-red color signals a would-be-terrorist and a macro-blue color means the person is not a would-be-terrorist. 

\begin{figure}[t]
\centering
\includegraphics[trim=0cm 2cm 0cm 3cm, clip, width=1\linewidth]{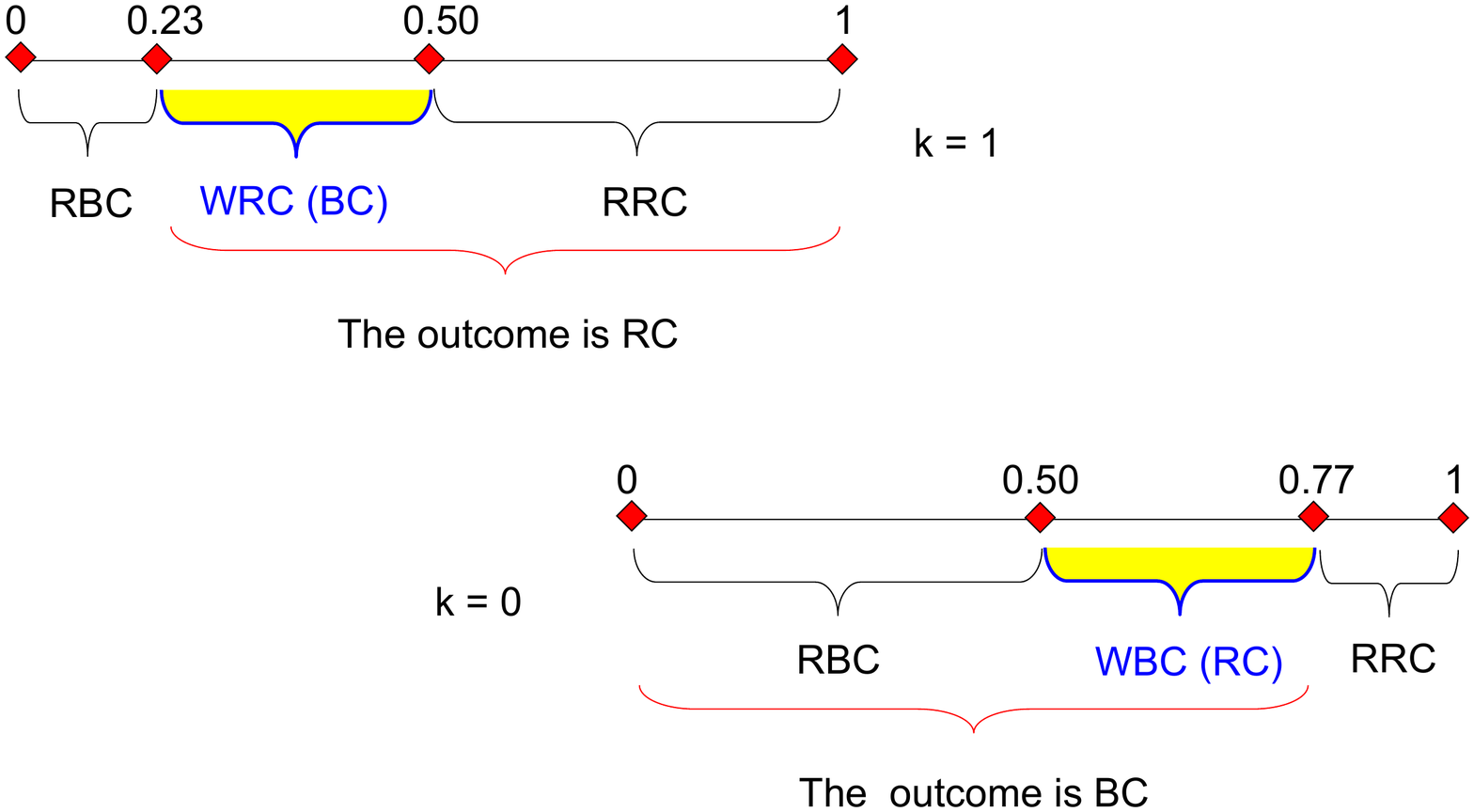}
\caption{The top part shows the macro-colors using $k=0$. The right B and R colors are respectively obtained for $0\leq p_0\leq 0.23$ and $0.50 \leq p_0\leq 1$ denoted RBC and RRC. However, a wrong R color (WRC) is found when $0.23 \leq p_0\leq 0.50$ instead of a B color (BC). The lower part shows the macro-colors using $k=1$. The right color (RBC) is obtained for $0\leq p_0\leq 0.50$. A wrong color (WBC) is found instead of the expected R color (RC) when $0.50 \leq p_0\leq 0.77$. The right color (RRC) is obtained for $0.77 \leq p_0\leq 1$.}
\label{rwc}
\end{figure}

At a tie, if the presumption of innocence prevails ($k=0$), a systematic error occurs for all persons whose scrutiny has yielded a proportion of terrorist compatible ground items (R) between $50\%$ and $77\%$. While these persons should be labeled would-be-terrorists they are wrongly declared as not would-be-terrorists as shown in the top part of Figure (\ref{rwc}). 

In contrast, applying a presumption of guilt ($k=1$) at ties ensures no would-be terrorists will be missed. But the concomitant price is that all persons with more than $23\%$ but less than $50\%$ of Terrorist Connected  ground items (R) will be wrongly tagged as would-be terrorists as seen in the lower part of Figure (\ref{rwc}) while indeed they are not would-be-terrorists.

\section{Adding a white third color for uninformative items}

In above two-color case, an ambiguous aggregate is given a property by coloring it either R or B depending on the current expectation of the monitoring of actual extraction of information.  However, to cover a larger spectrum of cases it is of interest to introduce a third color denoted white (W), to account for  uninformative items resulting from unclassified or ambiguous states, serving as a proxy for incomplete, missing, or uncertain data. White pixels can be present but event if they are not, W aggregates appear inevitably in the treatment of aggregates for which there exists no majority, either R, B, W.

The problem thus becomes a three color problem, R,B W, for which applying a majority rule to groups of four items generates unsolved configurations, for which none color is majority. Yet, a decision has to be made for each type of ambiguous configurations. The respective proportions of R, B, W  pixels are respectively $p_0, q_0, (1-p_0-q_0)$.

Distributing them randomly in groups of 4 can result in $3^4=81$ different configurations of which 15 have different compositions of R, B, W. Accounting fo the permutations in the respective configurations yields 3  with same color for the 4 pixels and 24 with the same color for 3 pixels. Applying majority rule to these 9 configurations yields 9 single-colored configurations as:

\begin{itemize}
\item RRRR (1), RRRB (4), RRRW (4)  $\Rightarrow$ RRRR
\item BBBB (1), RBBB (4), BBBW (4)$\Rightarrow$ BBBB
\item  WWWW (1), WWWR (4), WWWB (4) $\Rightarrow$ WWWW
\end{itemize}
where the numbers in parentheses are the numbers of equivalent configurations. 

For the remaining 6 configurations, no color has the majority, i.e., larger or equal to 3. Accordingly, these configurations required a special handling to decide which color to attribute to each one of them.

The ``synthesizer" must then select a series of criteria to allow a treatment of all cases. The associated rules can be directly related to a preconceived view on how to complete incomplete data or a random selection in tune with the framework culture of the system operating the data.

That is not a lack of rigor, on the contrary, to search for an insight required to be looking for some a priori expectation. This motivation will not reverse the synthesis when it is clear, here in the presence of a local majority, but when in doubt, here at a tie, this local a priori will make the difference, one way or the other.  An agent or an IA  sticks to the data when it is clear, but in presence of uncertainty, some a priori expectation is at work. That could be unconscious or conscious and is generally dictated by the corporate culture, and not by a conscious desire to manipulate the data. 

While the absence of a majority with only two colors in a group of four is self-evident, the situation is richer in the case of three colors. Absolute and relative majorities are now possible. In addition, the white, being without identified content, its relative weight is not equal to those of red and blue.

Accordingly, with W not counting, for a strong relative majority of 2 R (B) against 0 B (R), I take WWWW with a probability $u$ and $(1-u)$ for RRRR (BBBB). For a weak relative majority of 2 R (B) against 1 B (R), I take WWWW with a probability $v$ and $(1-v)$ for RRRR (BBBB). At a tie between R and B (2, 2; 1,1) I chose RRRR with probability $r$, BBBB with probability $b$ and WWWW with probability $(1-r-b)$. The following update rules apply to ambiguous configurations:

\begin{itemize}
\item RRWW $\Rightarrow$ WWWW, RRRR with respective probabilities $u, (1- u)$
\item BBWW $\Rightarrow$ WWWW, BBBB with respective probabilities $u, (1- u)$
\item RRBW $\Rightarrow$ WWWW, RRRR with respective probabilities $v, (1- v)$
\item RBBW $\Rightarrow$ WWWW, BBBB with respective probabilities $v, (1- v)$
\item RRBB, RBWW $\Rightarrow$ RRRR, BBBB, WWWW with respective probabilities $r, b, (1-r-b)$
\end{itemize}

Indeed, accounting for all different cases would require doubling the numbers of parameters from 5 to 10. I have chosen to restrict the investigation with the 5 parameters shown in Table (\ref{up}) to keep it focused and clearer without losing in generality. Given this choice, the update equations for one coarse-grained iteration from level-n to level-(n+1) write, 

\begin{align}
p_{n+1} & = p_n^2 \left[ p_n^2 + 4p_n q_n + 4p_n(1 - p_n - q_n) +6(1 - p_n - q_n)^2 (1 - u) \right. \notag \\
&\quad \left. + 12q_n(1 - p_n - q_n)(1 - v) \right]+b (6 p^2 q^2 + 12 p q (1 - p - q)^2), \label{p} \\ 
q_{n+1} & = q_n^2 \left[ q_n^2 + 4p_n q_n + 4q_n(1 - p_n - q_n) + 6(1 - p_n - q_n)^2 (1 - u) \right. \notag \\
&\quad \left. + 12p_n(1 - p_n - q_n)(1 - v) \right] +r (6 p^2 q^2 + 12 p q (1 - p - q)^2)  \label{q}  
\end{align}
with $(1-p_{n+1}-q_{n+1})$ yielding W for the aggregate.

Solving the associated fixed point equations $p_{n+1}=p_n$ and $q_{n+1}=q_n$ generates a 2-dimensional landscape for the dynamics of opinion instead of the previous 1-dimensional one with only the two colors R and B. Analytical solving is no longer feasible and a numerical treatment must be used. 

\begin{table}[h]
\centering
\renewcommand{\arraystretch}{1.4}
\begin{tabular}{>{\raggedright}p{7.5cm} >{\raggedright}p{5cm} >{\raggedright\arraybackslash}p{3cm}}
\toprule
\textbf{Inputs} & \textbf{Outputs} & \textbf{Condition} \\
\midrule
BBBB (1), BBBR (4), BBBW (4) & BBBB & $t_1 > 2$ \\
RRRR (1), RRRB (4), RRRW (4) & RRRR & $t_2 > 2$ \\
WWWW (1), WWWB (4), WWWR (4) & WWWW & $t_3 > 2$ \\
BBWW(6) & WWWW ($u$) \\ BBBB ($1 - u$) & $t_1 = t_3 = 2$ \\
RRWW(6) & WWWW ($u$) \\ RRRR ($1 - u$) & $t_2 = t_3 = 2$ \\
BBRW(12) & WWWW ($v$) \\ BBBB ($1 - v$) & $t_2 = t_3=1$ \\
RRBW(12) & WWWW ($v$) \\ RRRR ($1 - v$) & $t_1 = t_3 =1$ \\
BBRR (6), BRWW(12)  &  BBBB ($b$) \\ RRRR ($r$) \\ WWWW ($1 - b-r$) & $t_1 = t_2 = 2, 1$ \\
\bottomrule
\end{tabular}
\caption{Update rules with multiline probabilistic outputs with $t_1=$ number of B, $t_2=$ number of R and $t_3=$ number of W. The numbers in parentheses signals the numbers of equivalent configurations obtained by permuting the colors. The letters in parentheses are the probabilities of the respective outcomes of the updates.}
\label{up}
\end{table}

\section{Results from the update equations}

My main focus in this work is not to discuss the nature and merits of each choice of treatment of the various ambiguities implemented by $(u, v ,r, b)$. The focus is to demonstrate that the appearance of ambiguities is inevitable from the coarse-graining process, and that no matter how they are sorted, some pixel compositions $(p, q)$ will result in erroneous macro-color diagnoses.

\begin{figure}
\includegraphics[width=1\textwidth]{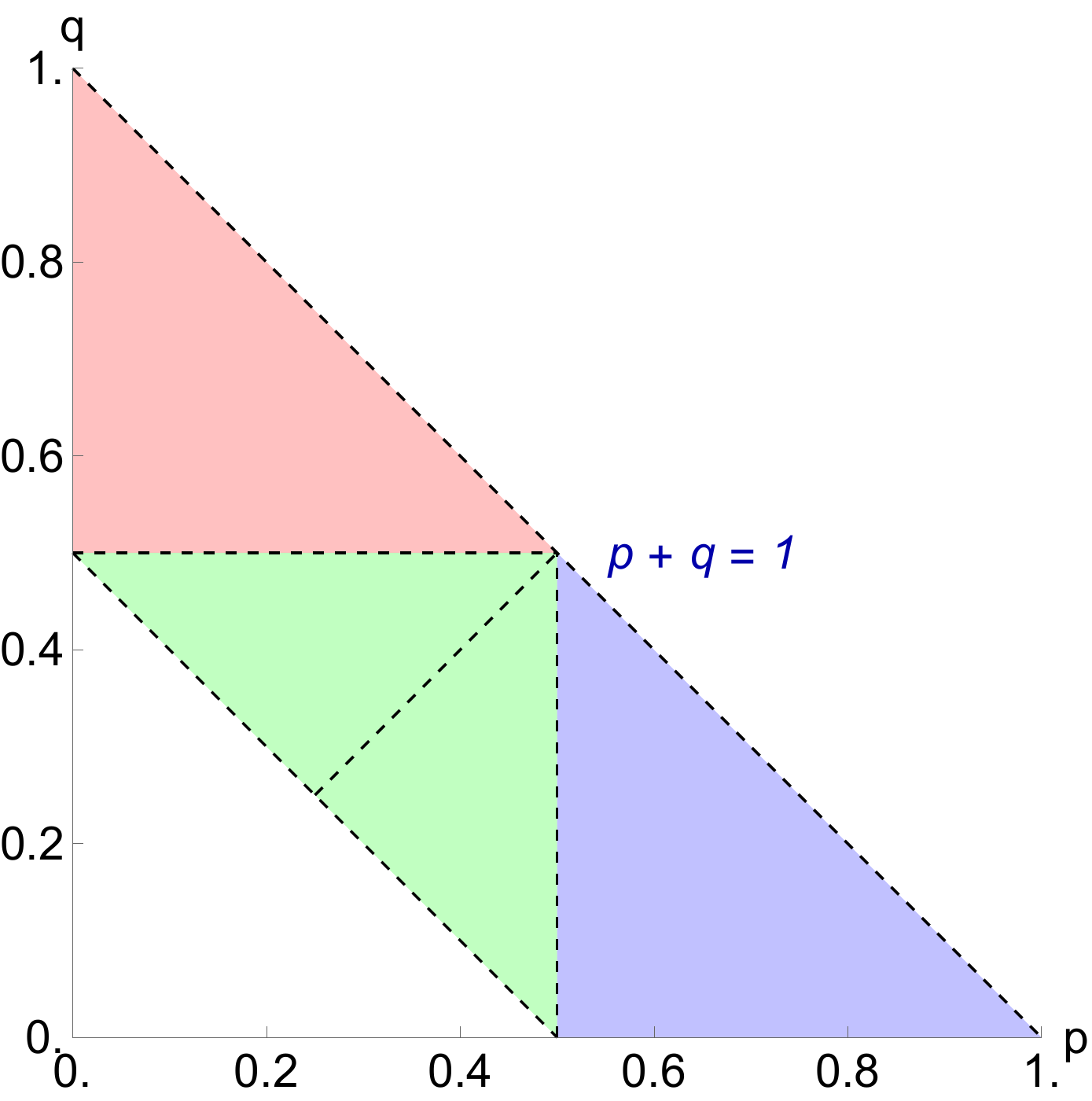}
\caption{The 2-dimensional $(p, q)$ diagram for all possible compositions of B, R, W pixels and aggregates. The area in blue (red, white) has more than fifty percent of B (R, W) making B (R, W) the associated macro-color. The area in green has no colored absolute majority but combined B and R are more than fifty percent. Collections of pixels without W are located on the line $p+q=1$. }
\label{f0}
\end{figure}

On this basis, the parameter space being of six dimensions $(p, q, u, v, r, b)$, I restrict the investigation to a series of representative cases without losing in generality. The underlying surface of expected exact macro-colors is exhibited in Fig (\ref{f0}) as a function of  distributions of $p$ and $q$. The light blue, red and white triangles delimit the areas with a majority of pixels, respectively B, R, W. There, exact macro-colors are B, R, W.  In the central green triangle no absolute majority (more than half) prevails but $p+q > 0.50$ with a relative majority of one of the two colors in relation to the other. An arbitrage is thus required to set the related expected macro-color if any.

For each set of chosen  parameters $(u, v, r, b)$ I identify all associated fixed points using Eqs. (\ref{p}, \ref{q}) together with their respective stabilities. I then build the related 2-dimensional complete flow diagram generated by the coarse-graining from every point $(p, q,)$ till its ending attractor. Stable, unstable, saddle  fixed points are shown respectively in blue, red, magenta in the Figures.

The obtained diagram indicates areas of pixel composition ending up in wrong macro-color outcomes. As seen from Table (\ref{up}) I consider a symmetry of B and R against W using $u$ for both BBWW and RRWW configurations and $v$ for both BBRW and RRBW. At this stage including an B - R asymmetry would only blur the readability of the results. Assuming full symmetry between R and B implies $r=b$.

It is worth emphasizing that $p+q<1$ implies a proportion $(1-p-q)$ of white pixels in the sample. In addition, white aggregates appear during the coarse-graining implementation as a function of the treatment of local ambiguities depending on the value of $(u, v, r, b)$. With this regard, for each $(u, v, r, b)$ set, to discriminate the impact of white pixels from the forming of white aggregates, I show two flow diagrams. One including the full two-dimensional triangular surface, which embodies any pixel composition $(p,q, 1-p-q)$  and the one with samples with no white pixels, i.e., along the one-dimensional line $p+q=1$.

\begin{figure}[htbp]
\centering
\makebox[\textwidth][c]{
\includegraphics[width=0.65\textwidth]{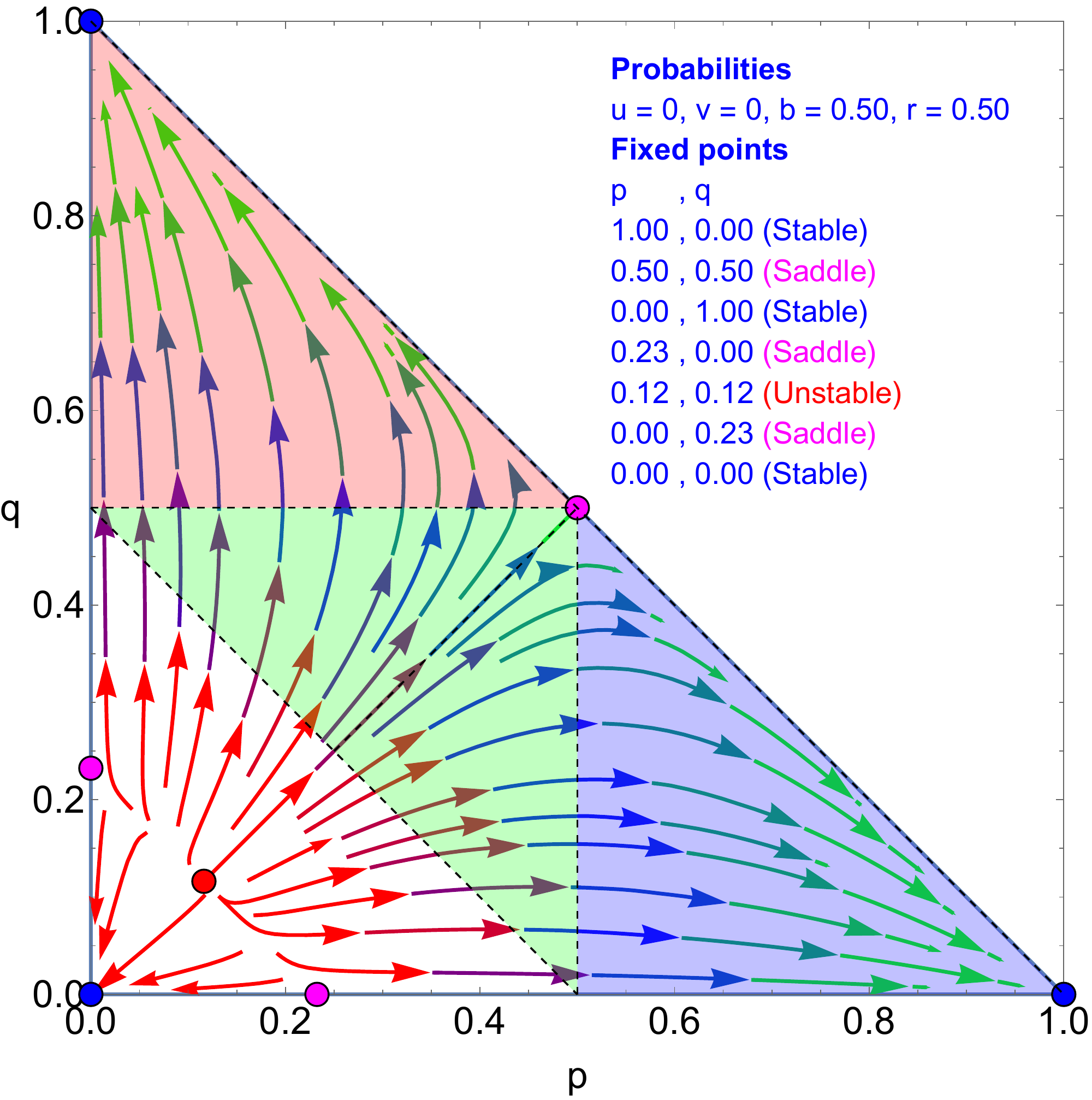}
\includegraphics[width=0.65\textwidth]{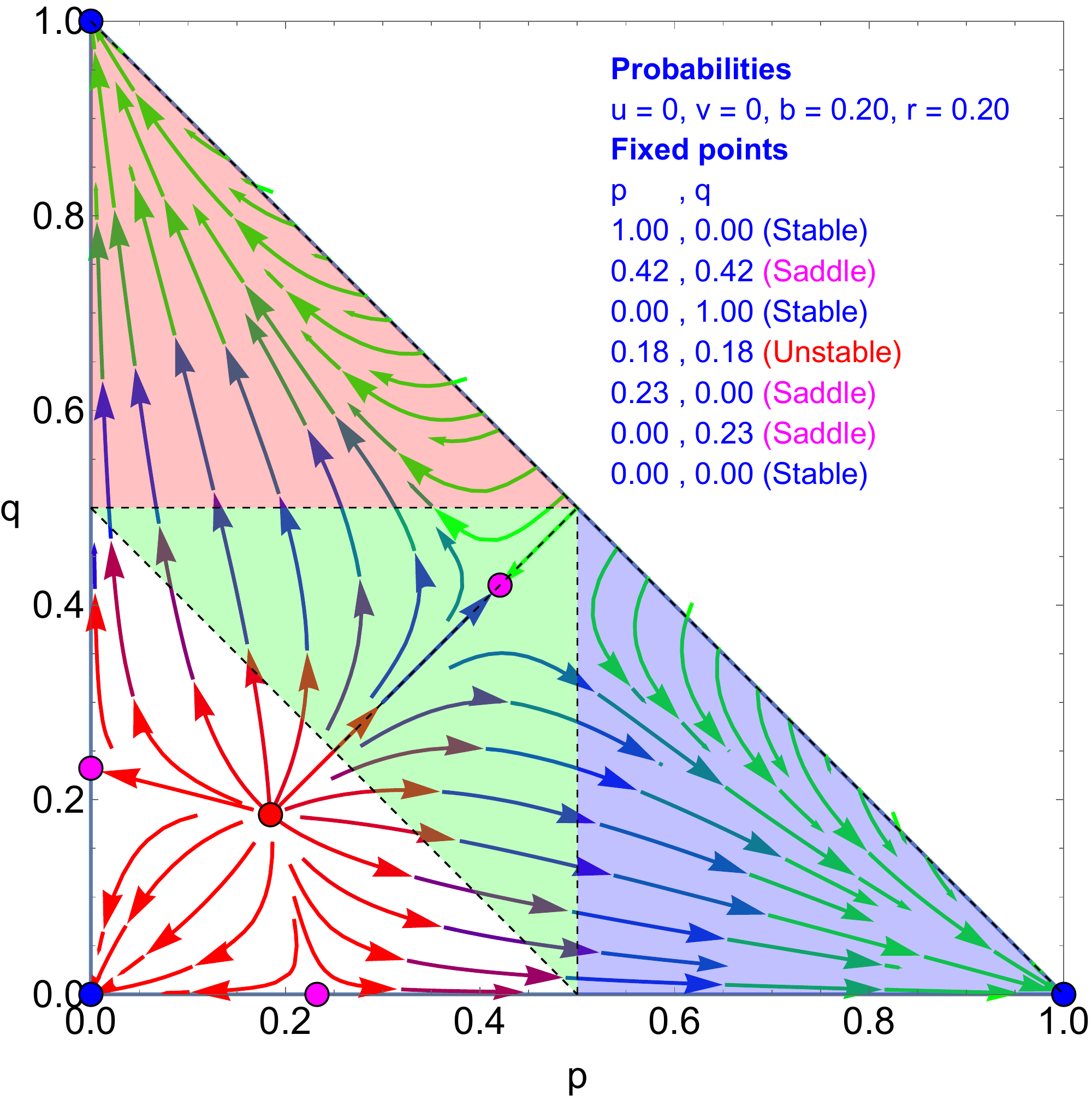}
}
\vspace{0.5cm}
\makebox[\textwidth][c]{
\includegraphics[width=0.65\textwidth]{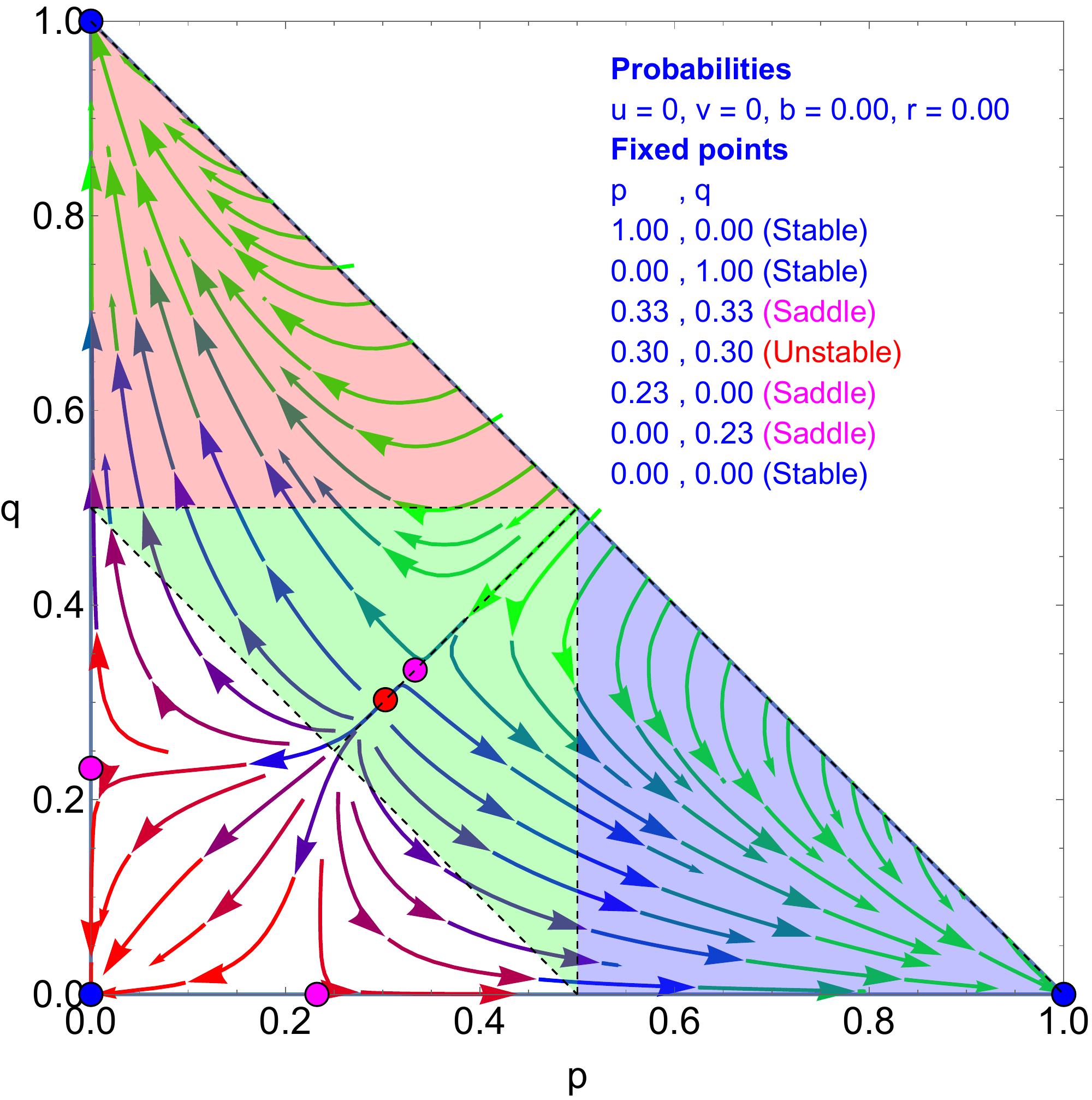}
\includegraphics[width=0.65\textwidth]{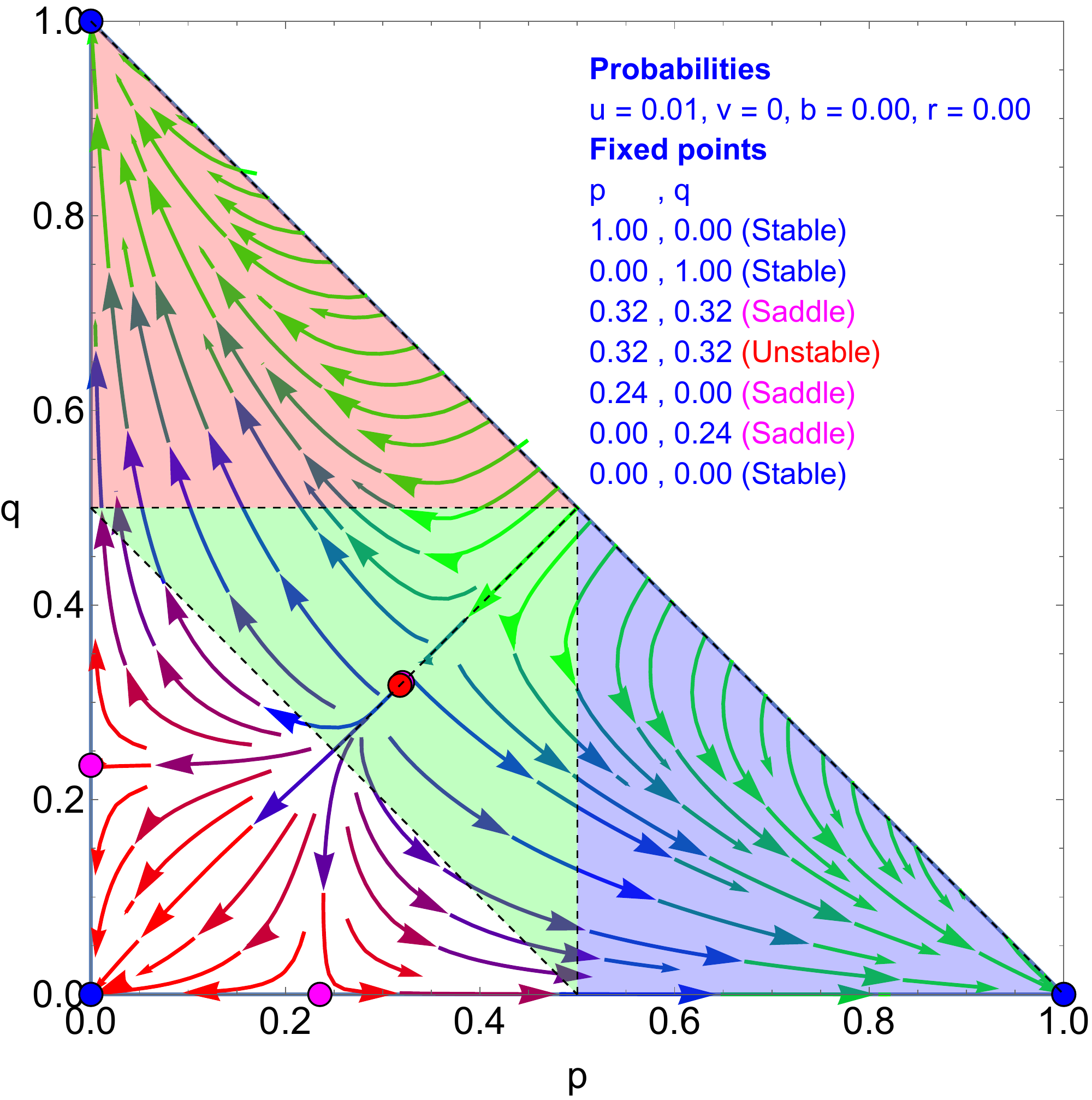}}
\caption{The full flow diagram generated by the coarse-graining for for sets of parameters $u=v=0, b=r=0.50$ (upper left), $u=v=0, b=r=0.20$ (upper right), $u=v=0, b=r=0$ (lower left), $u=0.01, v=0, b=r=0$ (lower left). For each case all associated fixed points are listed together with their respective stabilities. They are shown in the Figure with colored disks.}
\label{f14}
\end{figure}

I start with W does not counting in the local calculations of the majority, i.e., $u=v=0$. Thus, W disappears always besides for WWWB and WWWR, which yields WWWW following majority rule. If W does not count locally, $b=r=0.50$. It means that there is no tie breaking effect. Seven fixed points are found and listed in the upper left part of Fig. (\ref{f14}). The bassin of attraction of W attractor ($p=q=0$) is at its minimal area. The green area leads equally to B ($p=1, q=0$) and R ($p=0, q=1$) attractors.

When BBRR and BRWW yield WWWW with a probability $(1-b-r) \neq 0$,  $b=r< 0.50$.
The case $b=r=0.20$ is shown in the upper right part of Fig. (\ref{f14}). Comparing with the upper left part indicates that the flow along the $q=0$ and $p=0$ lines are unchanged with a tipping point still located on each at 0.23. However, the two unstable and saddle fixed points ($p=q=0.12$) and ($p=q=0.50$) have moved toward each other with $p=q=0.18$ and $p=q=0.42$.

With ($b=r=0$) the two unstable and saddle fixed points move to ($p=q=0.3$) and ($p=q=0.33$) but do not overlap as seen in the lower left part of Fig. (\ref{f14}). Only $u=0.01$ generates the overlap at ($p=q=0.32$) as shown in the lower right part of Fig. (\ref{f14}). In addition the tipping points on the axes move to 0.24. 

\begin{figure}[htbp]
\centering
\makebox[\textwidth][c]{
\includegraphics[width=0.65\textwidth]{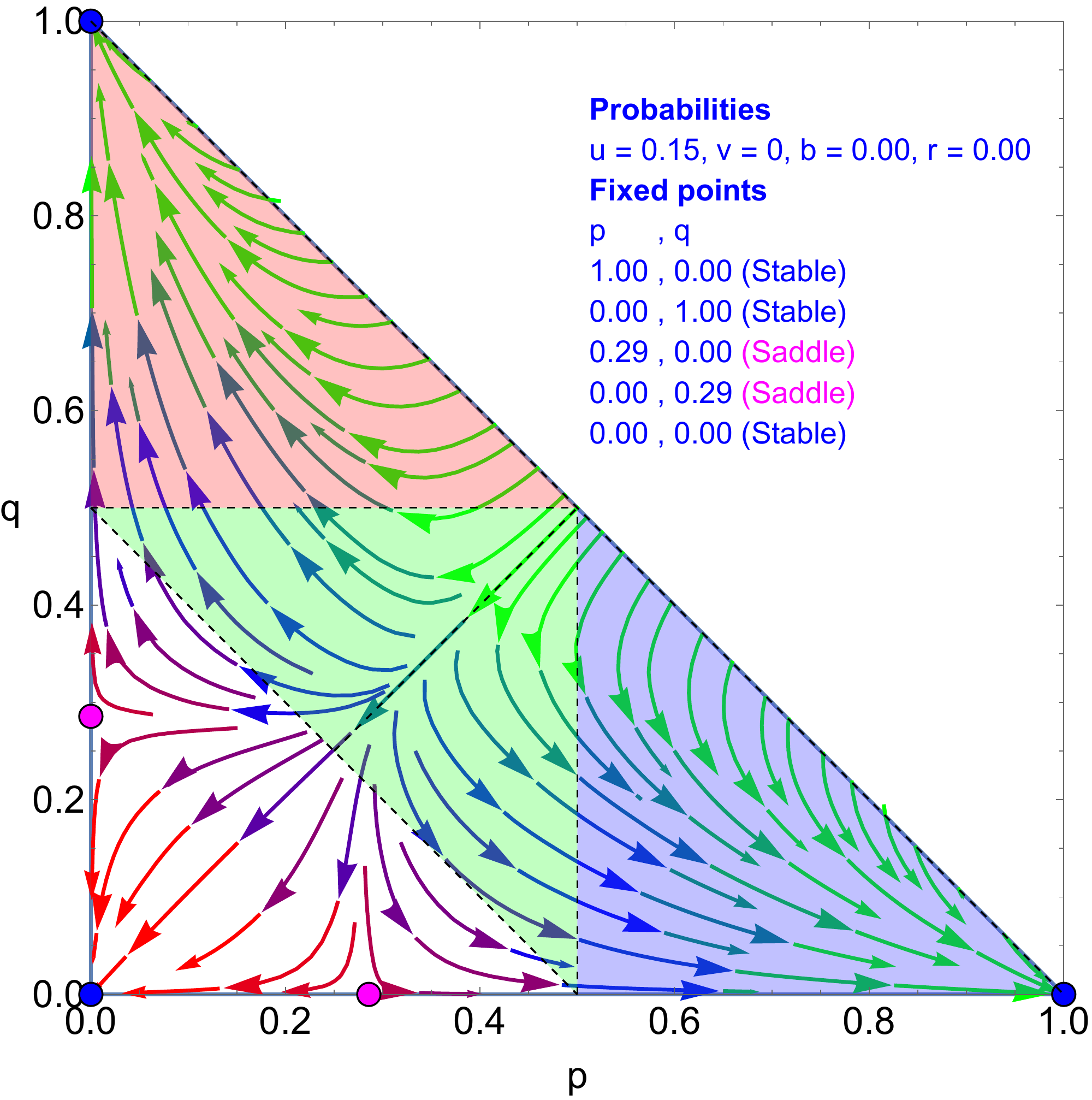}
\includegraphics[width=0.65\textwidth]{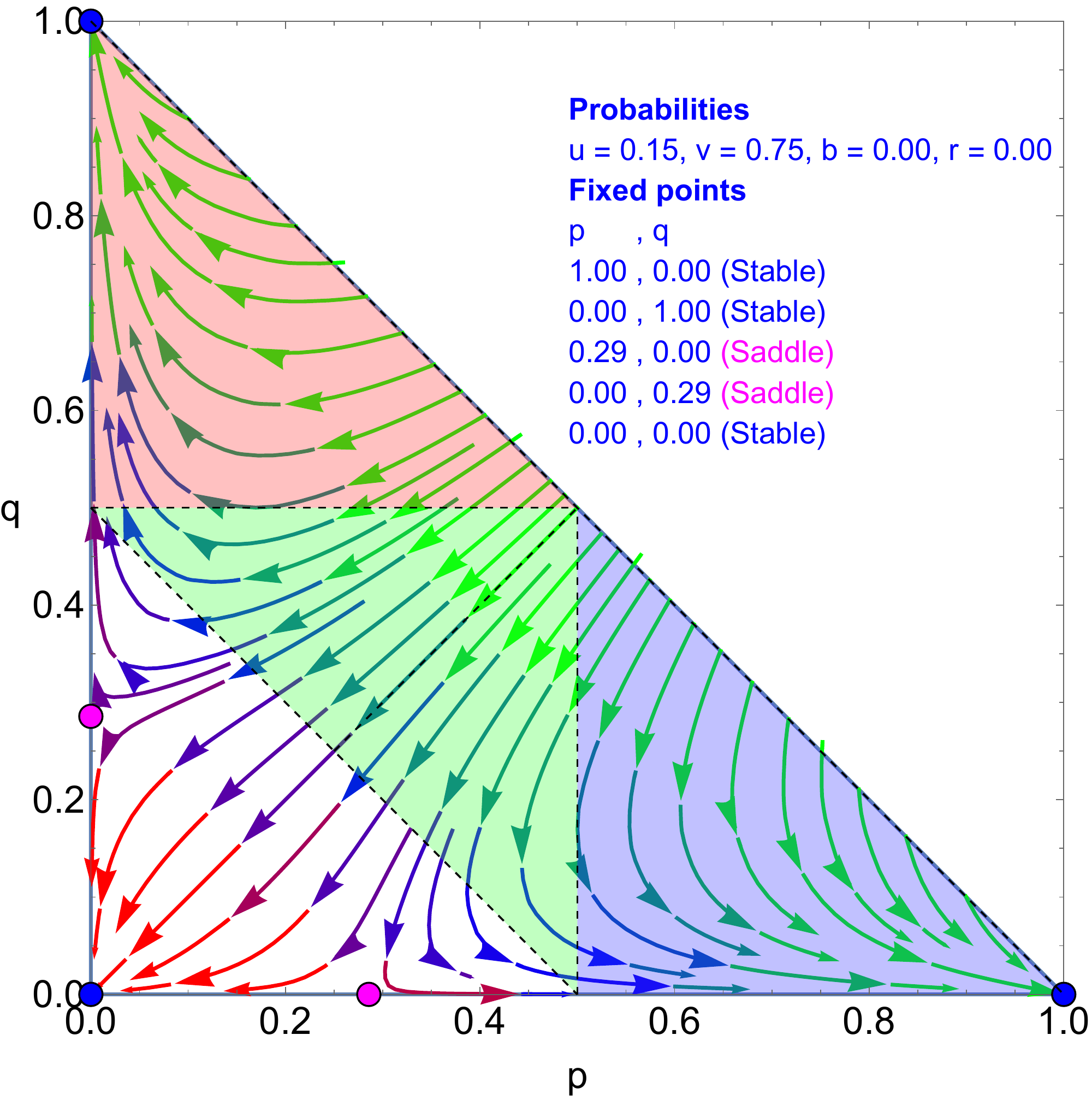}
}
\vspace{0.5cm}
\makebox[\textwidth][c]{
\includegraphics[width=0.65\textwidth]{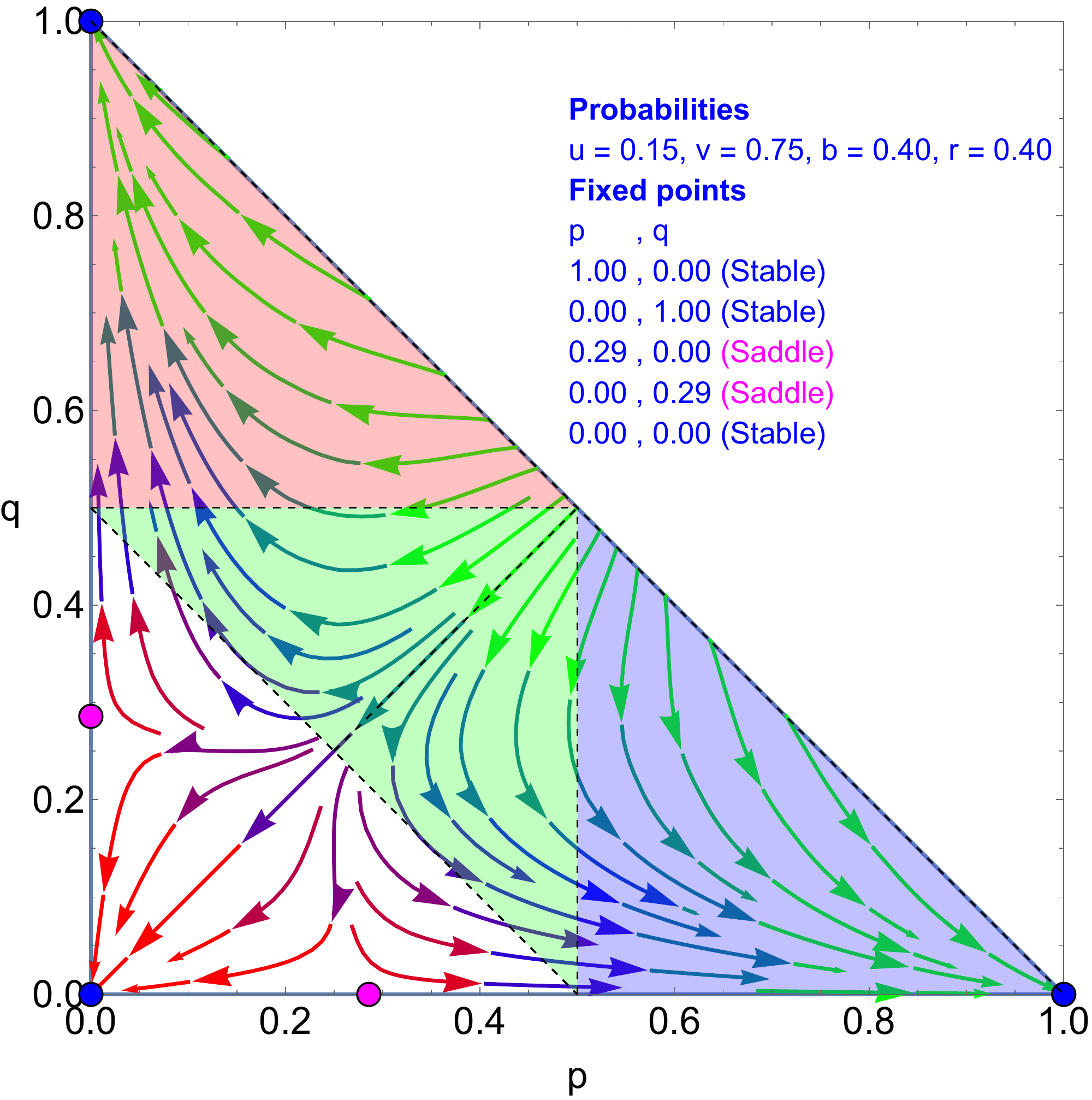}
\includegraphics[width=0.65\textwidth]{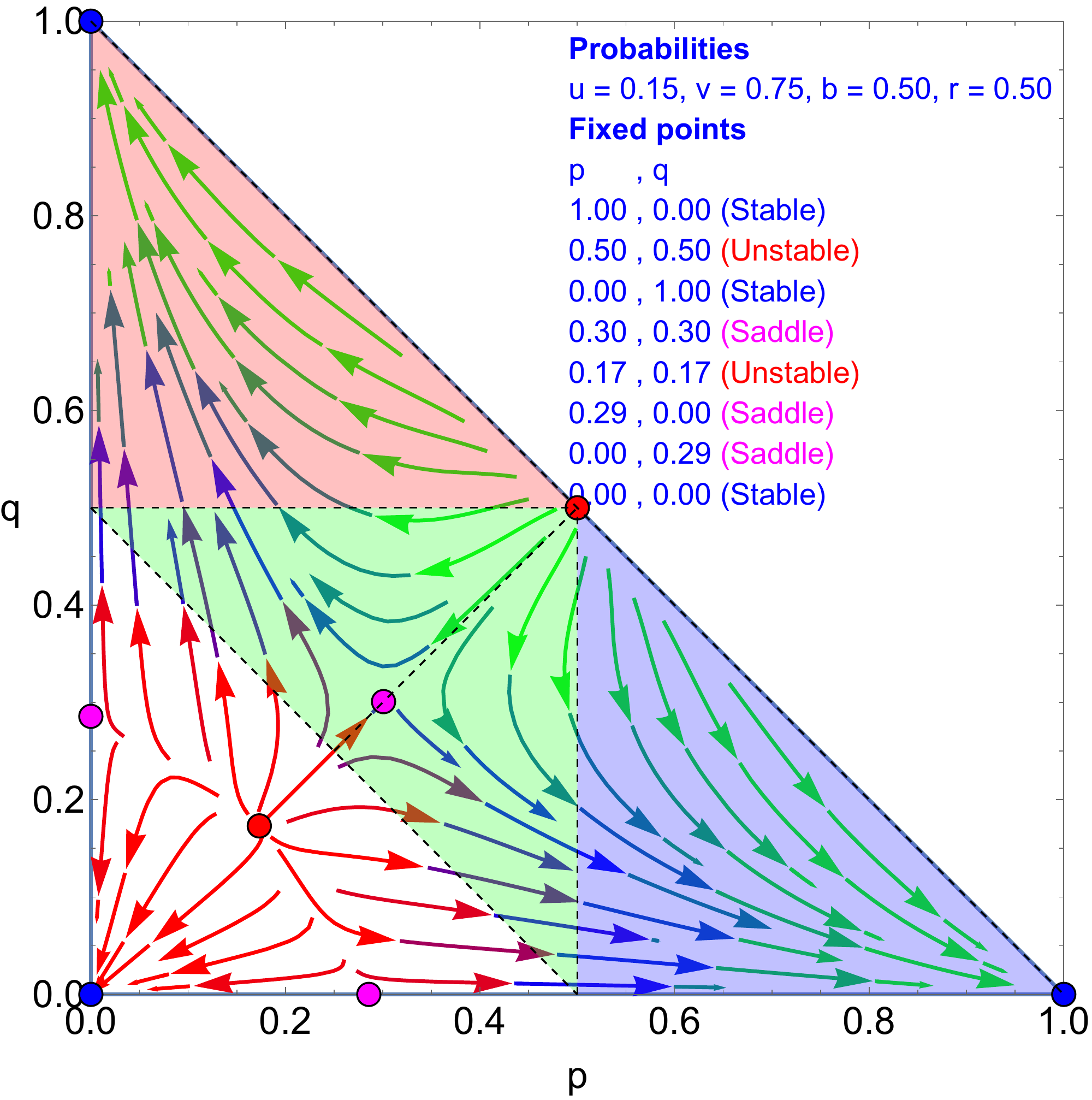}}
\caption{The full flow diagram generated by the coarse-graining for for sets of parameters $u=0.15, v=0, b=r=0$ (upper left), $u=0.15, v=0.75, b=r=0$ (upper right), $u=0.15, v=0.75, b=r=0.40$ (lower left), $u=0.15, v=0.75, b=r=0.50$ (lower left). For each case all associated fixed points are listed together with their respective stabilities. They are shown in the Figure with colored disks.}
\label{f58}
\end{figure}

Once they coalesce the fixed points ($p=q=0.32$) disappear leaving the flow landscape driven with only 5 fixed points as illustrated in the upper left part of Fig. (\ref{f58}) for $u=0.15$. Moving to the case $u=0.15, v=0.75, b=r=0$ has no much effect on the dynamics, besides directing the flows more quickly toward the two attractors $(1,0), (0,1)$ as exhibited in the upper right part of Fig. (\ref{f58}). Increasing $(b,r)$ from $(0,0)$ to $(0.40)$ has not much of effect as shown in the lower left part of Fig. (\ref{f58}). However, $(b=r=0.50)$ reintroduces 3 fixed points to reach a total of 7 as seen in the lower right part of Fig. (\ref{f58}).

\begin{figure}[htbp]
\centering
\makebox[\textwidth][c]{
\includegraphics[width=0.65\textwidth]{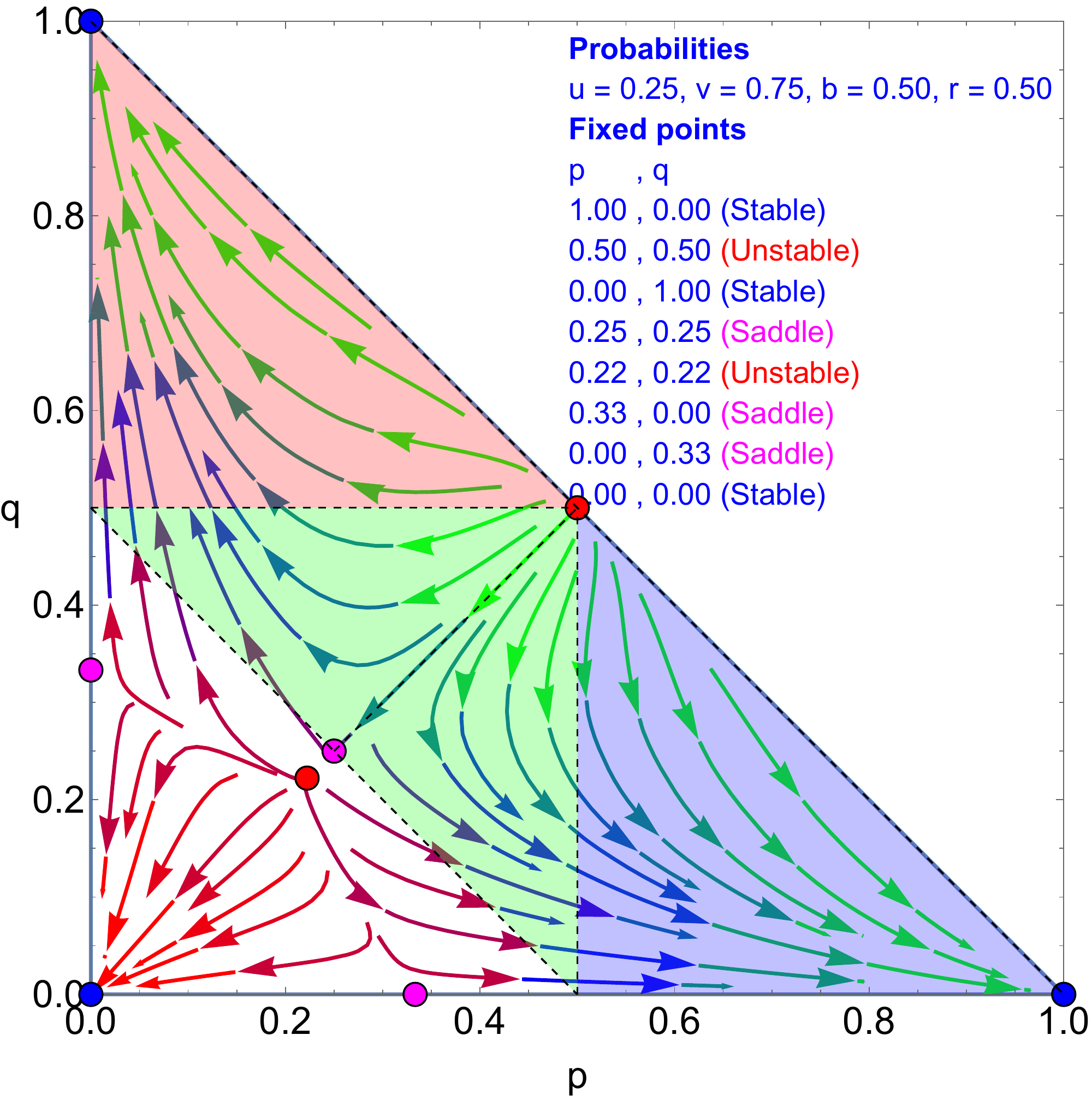}
\includegraphics[width=0.65\textwidth]{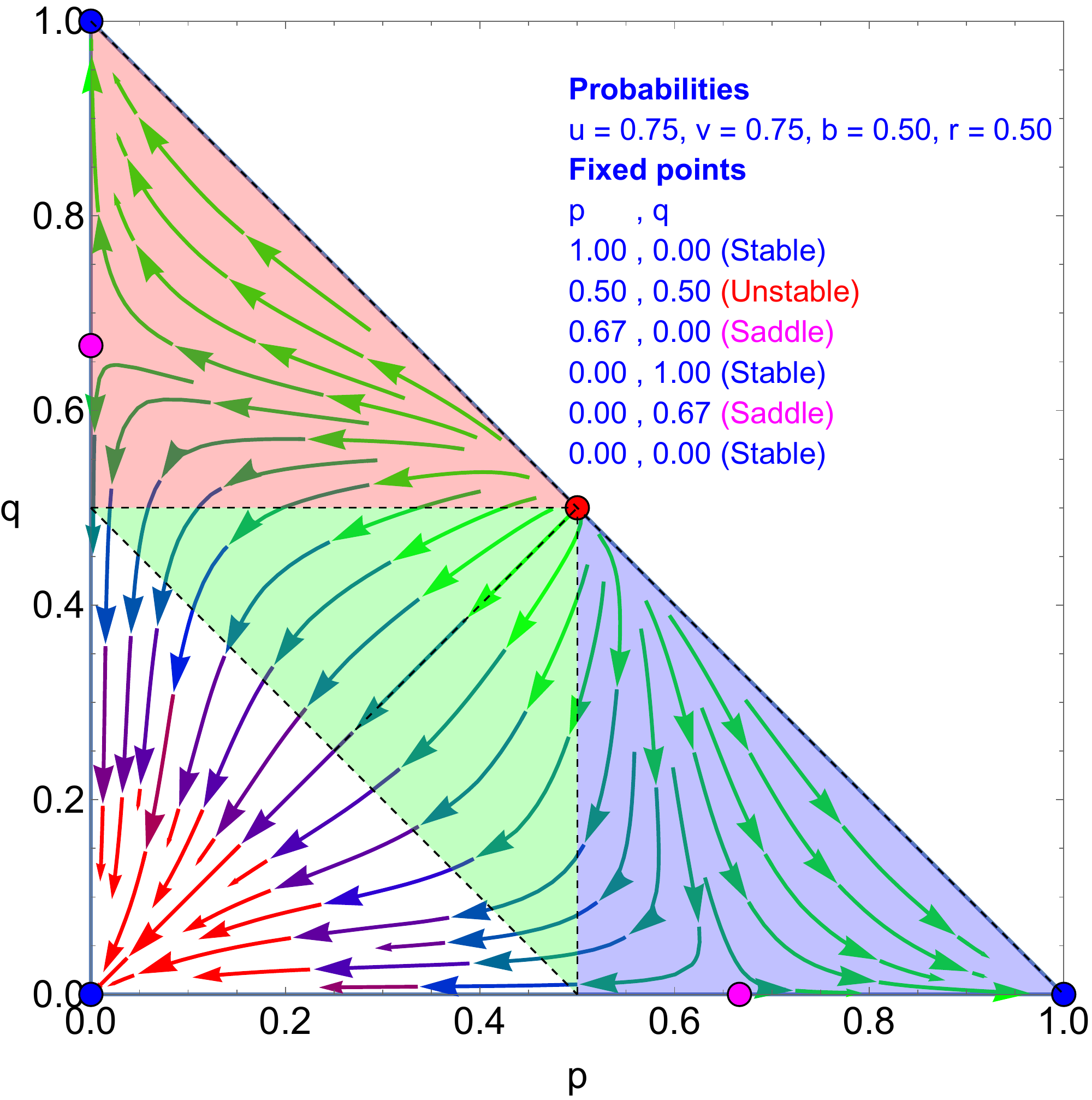}
}
\vspace{0.5cm}
\makebox[\textwidth][c]{
\includegraphics[width=0.65\textwidth]{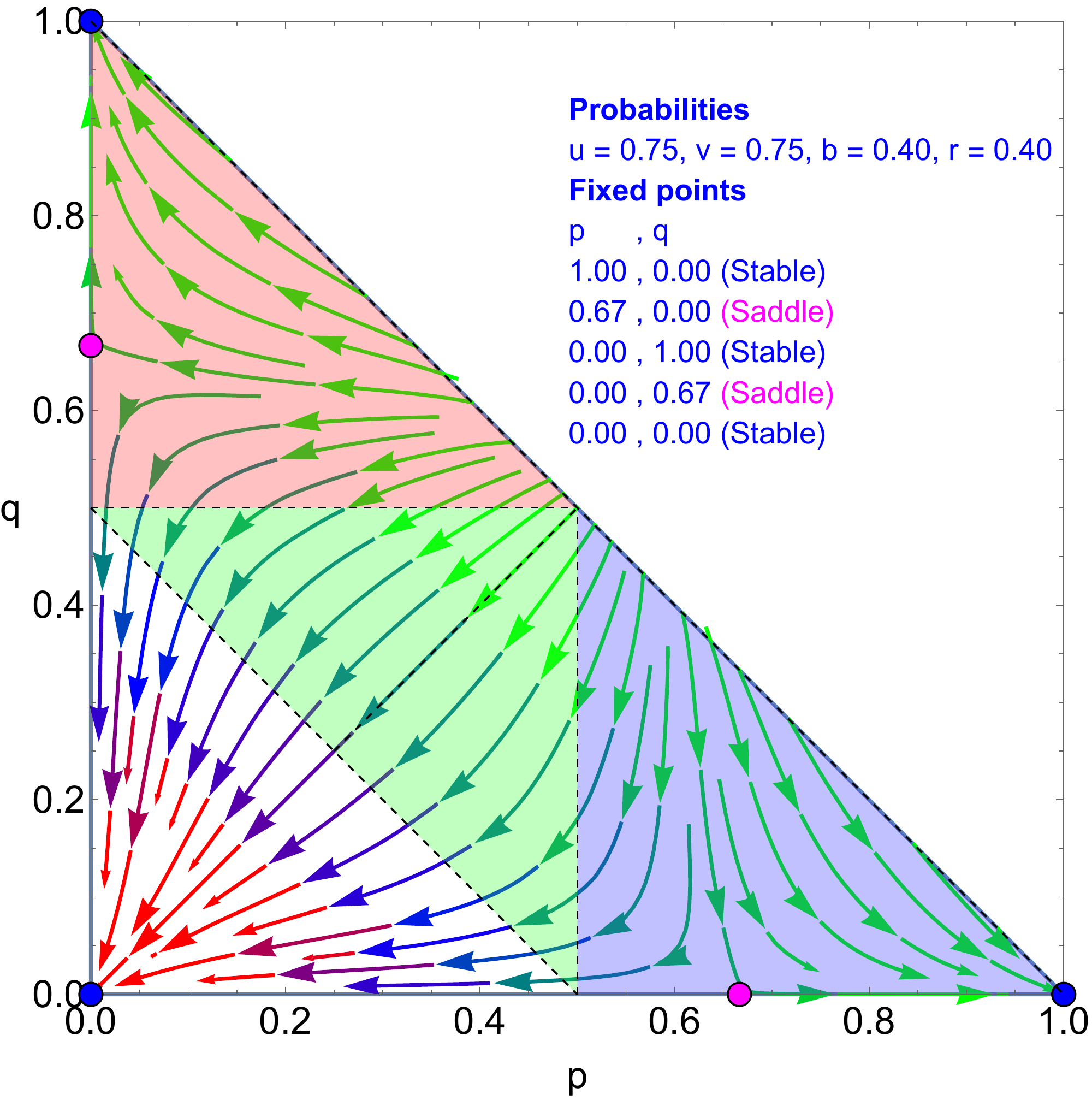}
\includegraphics[width=0.65\textwidth]{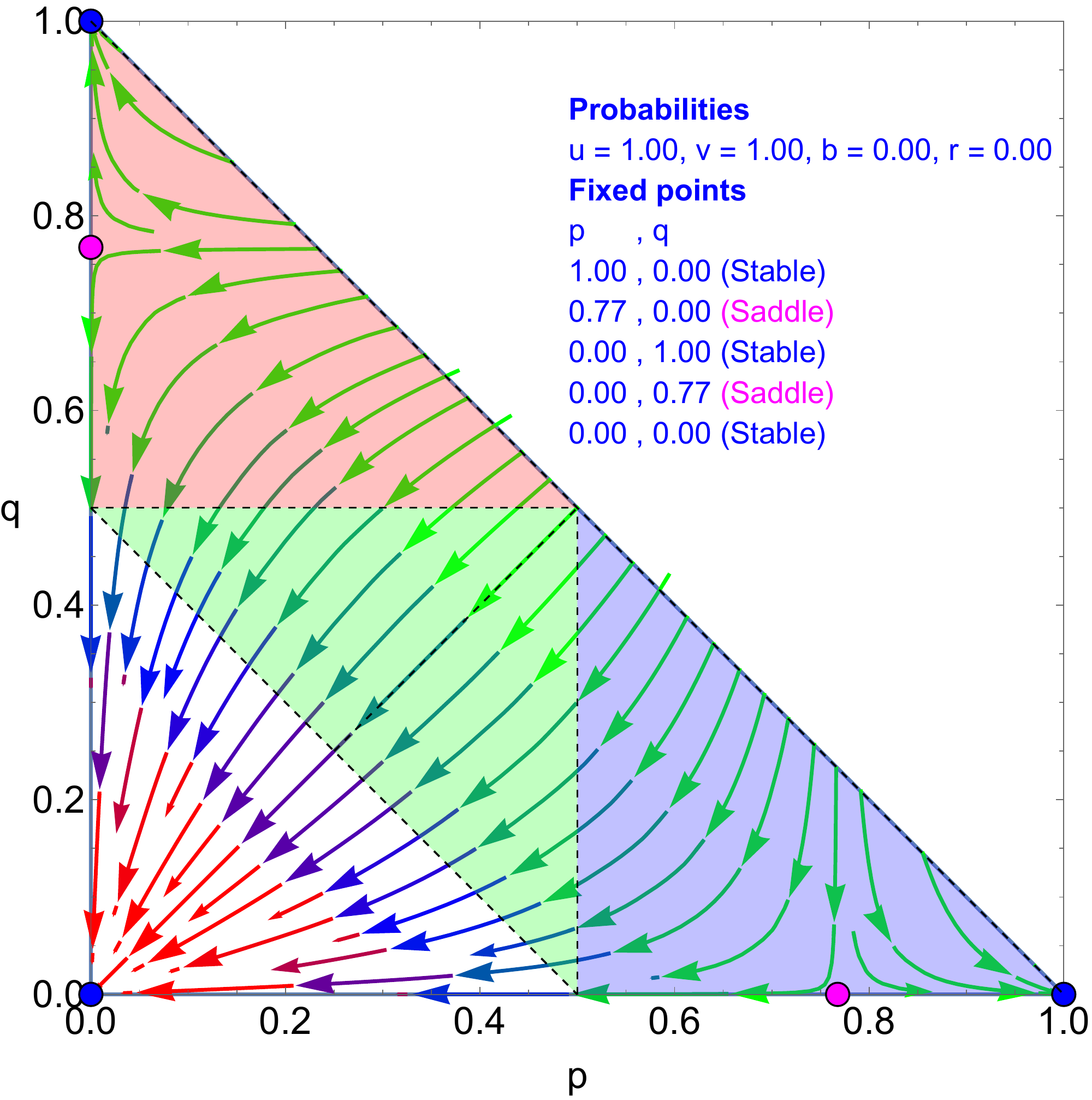}}
\caption{The full flow diagram generated by the coarse-graining for for sets of parameters $u=0.25, v=0.75, b=r=0.50$ (upper left), $u=v=0.75, b=r=0.50$ (upper right), $u=v=0.75, b=r=0.40$ (lower left), $u=v=1, b=r=0$ (lower left). For each case all associated fixed points are listed together with their respective stabilities. They are shown in the Figure with colored disks.}
\label{f812}
\end{figure}

Increasing $u$ from 0.15 to 0.25 keeping unchanged $v=0.75, b=r=0.50$ has not much effect. But for $u=0.75)$ two fixed points disappeared leaving the dynamics driven by 6 fixed points as exhibited respectively in the left and right part of the upper part of Fig (\ref{f812}). The lower part of Fig (\ref{f812}) shows that while $(u=v=0. 75, b=r=0.40)$ has not much effect, $(u=v=1, b=r=0)$ shifts the axis attractors from 0.67 to 0.77.

All above case have enlightened the unavoidable flaws produced by the handling of local ambiguities in the way they are treated contributing in part to B, R and W.

\section{Results from simulations}

To validate above results I run simulations to materialize the actual coarse-graining of a collection of colored pixels. Since here one simulation is required for each pair $(p_0, q_0)$, I am reporting only three cases to illustrate the process. I treat collections of 1064 pixels randomly distributed on a $2^5 \times 2^5 = 64  \times 64$ grid, which corresponds to five consecutive updates. Each level of the coarse-graining is numbered $l= 1, 2, 3, 4, 5$ with $l=0$ for the actual collection of pixels. For each level $l$, $t_1, t_2, t_3$ denote the  proportions of aggregates respectively blue, red, white. Parameters $t_1, t_2, t_3, l$ are identical to $p_n, q_n, 1-p_n-q_n, n$ used for above update equations.

First case is associated to point $t_1=0.14, t_2=0.11, t_3=0.75$ given $u=v=0$, $b=r=0.50$. The upper left part of Fig. (\ref{f14}) lists the associated seven fixed points driven the dynamic flow.  In particular, starting from $p_0=0.14, q_0=0.11$ leads to the W attractor $p_0=q_0=0$, which is the correct label $75\%$ of W pixels. The related five levels are exhibited in Fig. (\ref{a05}) with the same macro-color.

I noticed that $t_1=0.14, t_2=0.11$ is within the basin of attraction delimited by the unstable fixed point (0.12, 0.12) and two saddle points at (0.23, 0) and (0, 0.23).

\begin{figure}[htbp]
\vspace{-3cm}
\centering
\makebox[\textwidth][c]{
\includegraphics[width=0.45\textwidth]{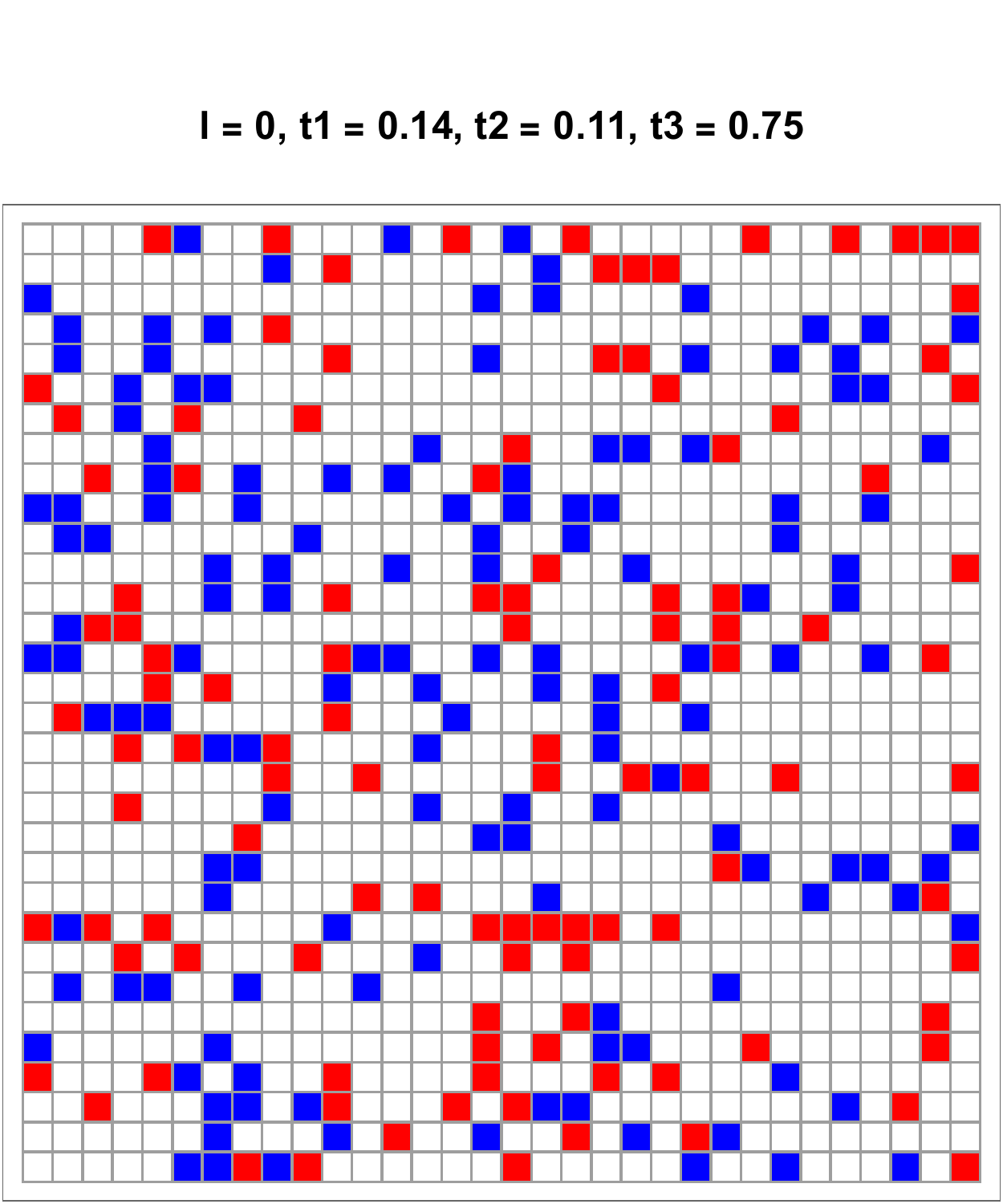}
\includegraphics[width=0.45\textwidth]{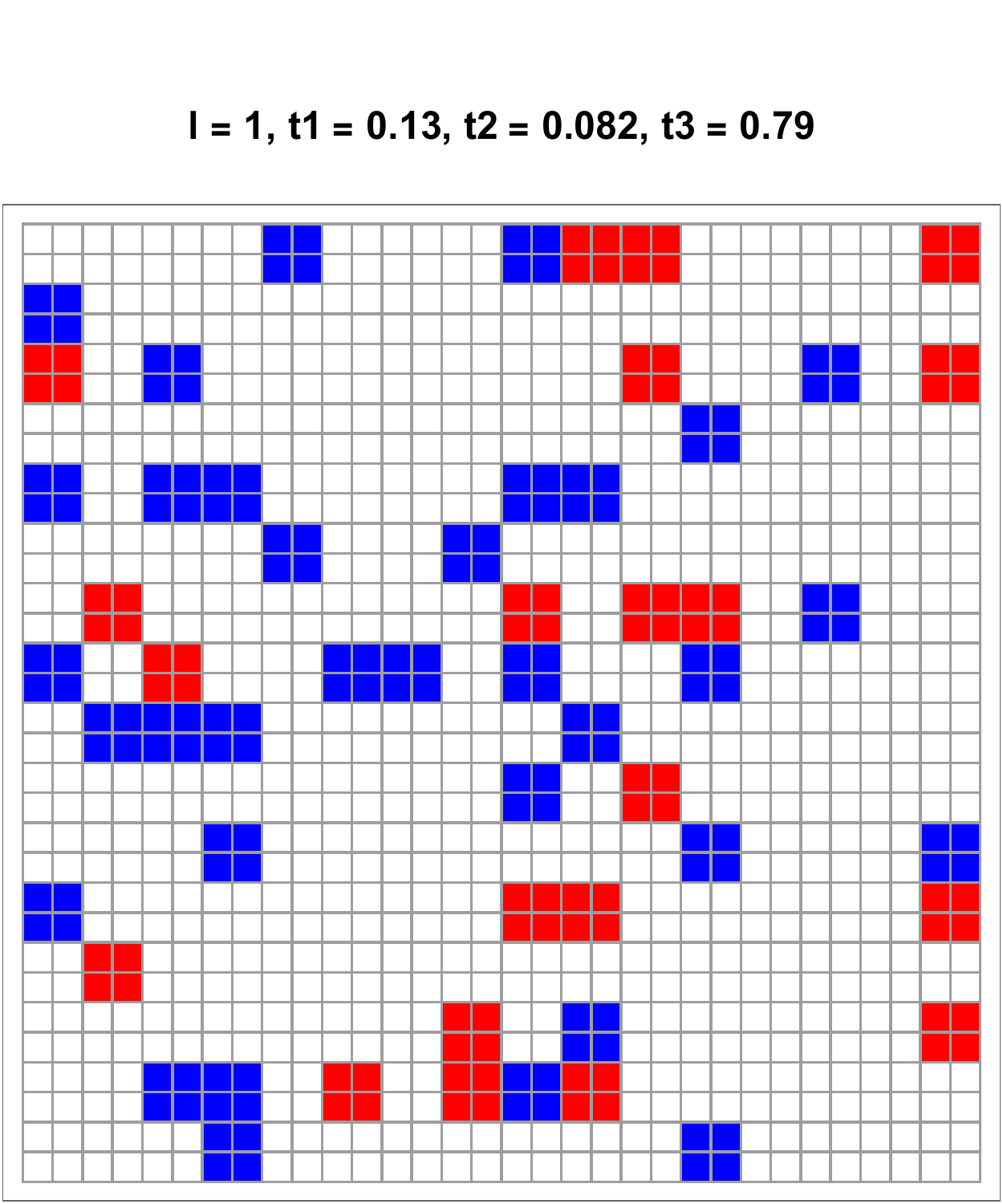}}
\vspace{0.5cm}
\makebox[\textwidth][c]{
\includegraphics[width=0.45\textwidth]{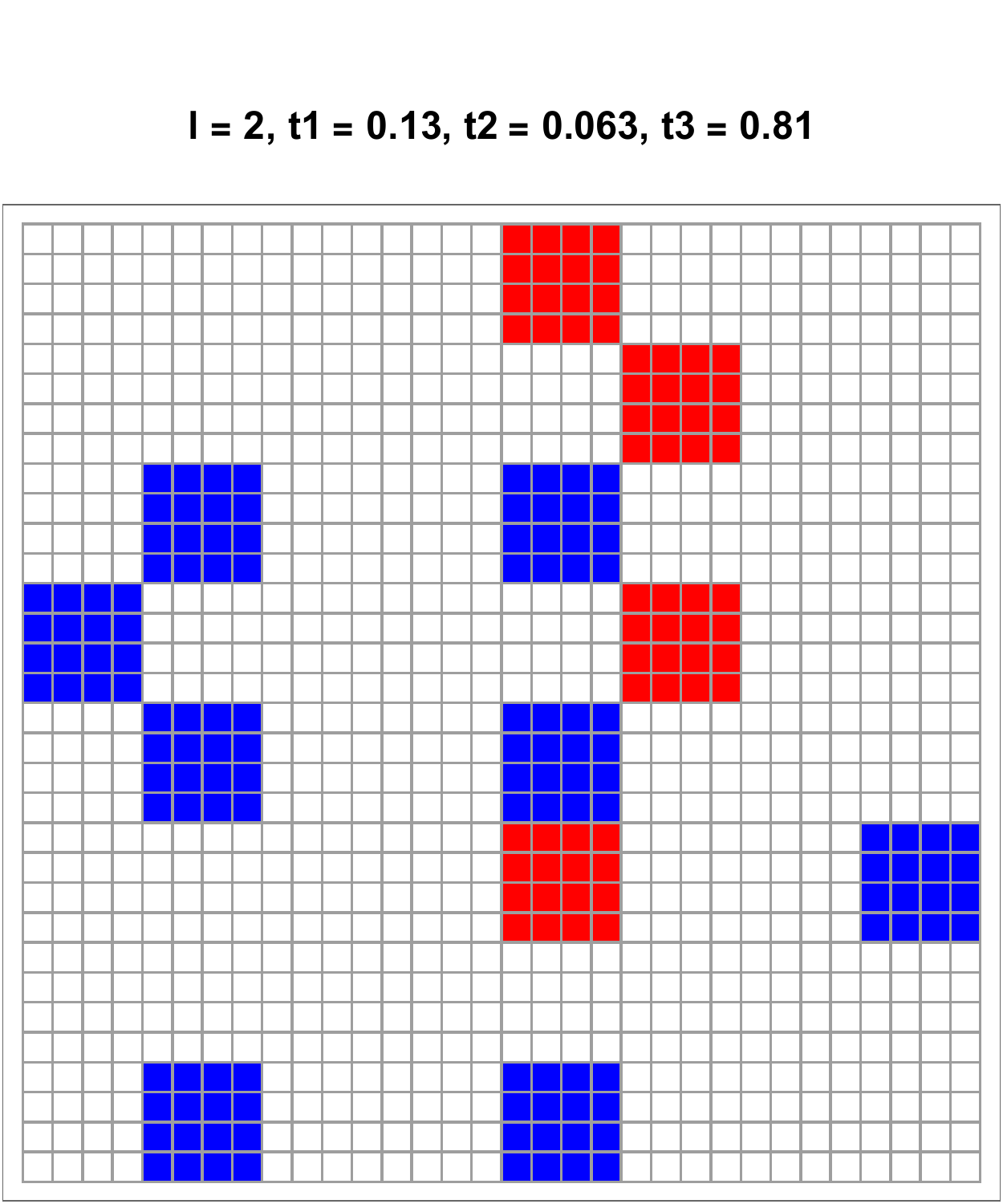}
\includegraphics[width=0.45\textwidth]{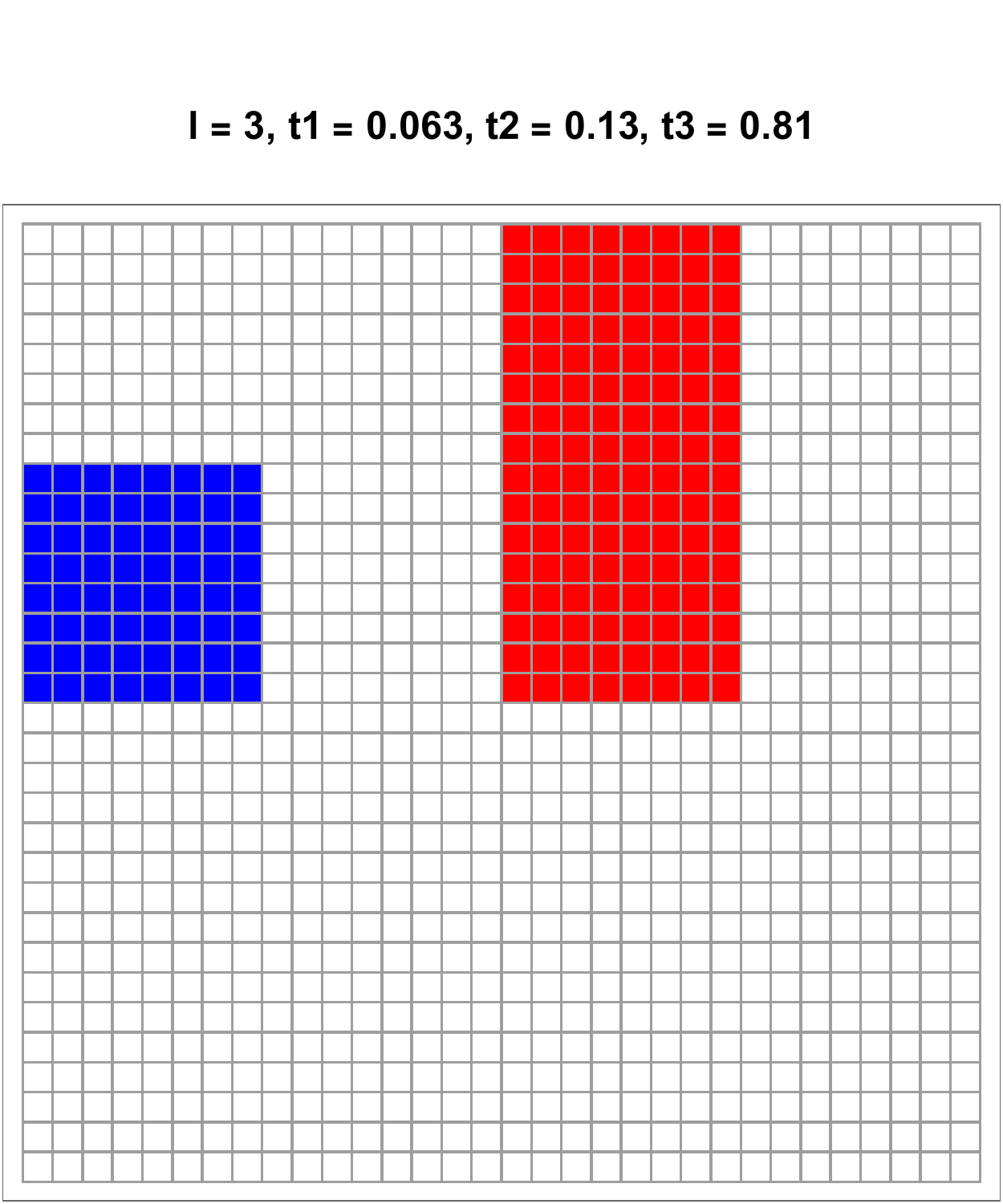}}
\vspace{0.5cm}
\makebox[\textwidth][c]{
\includegraphics[width=0.45\textwidth]{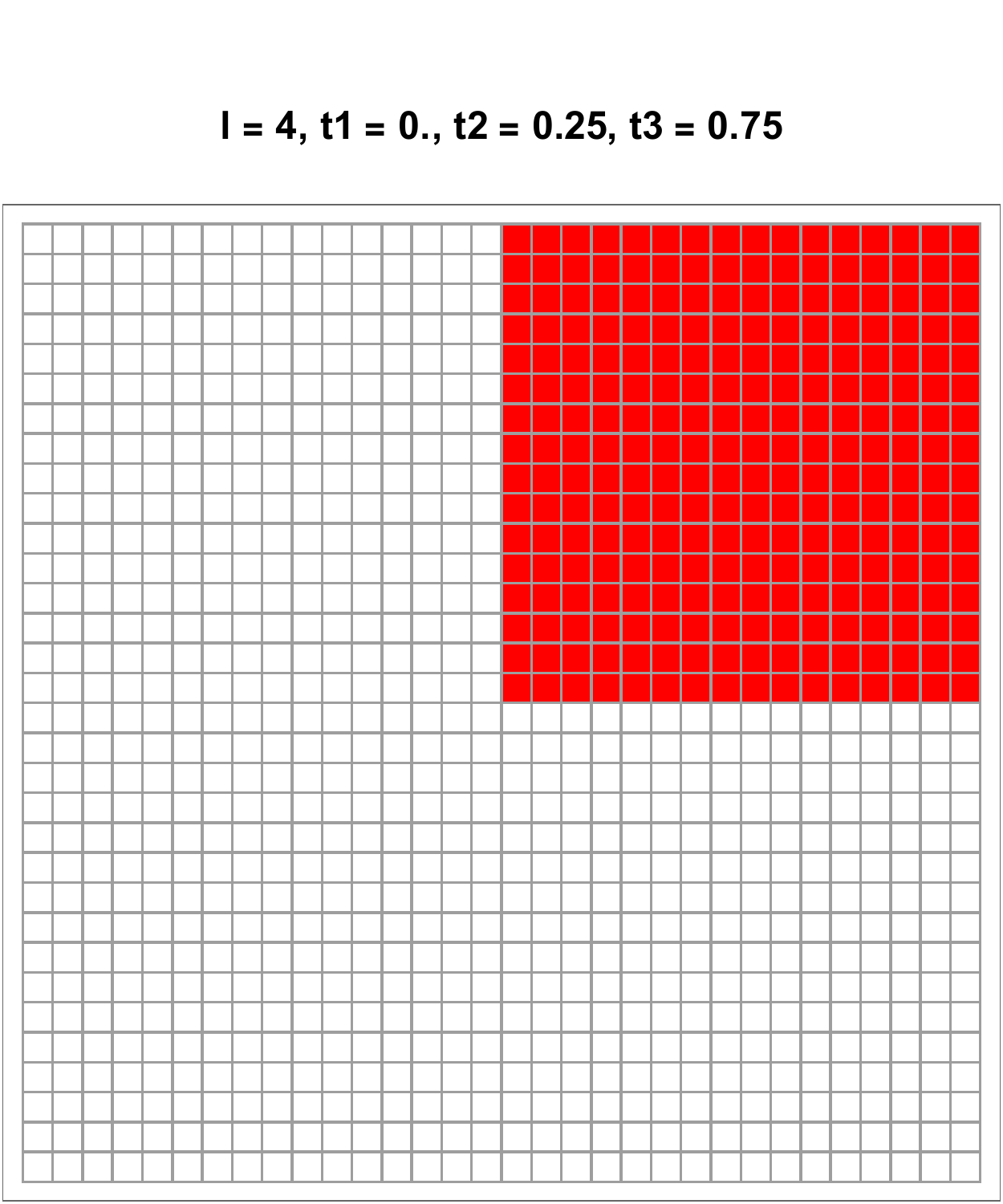}
\includegraphics[width=0.45\textwidth]{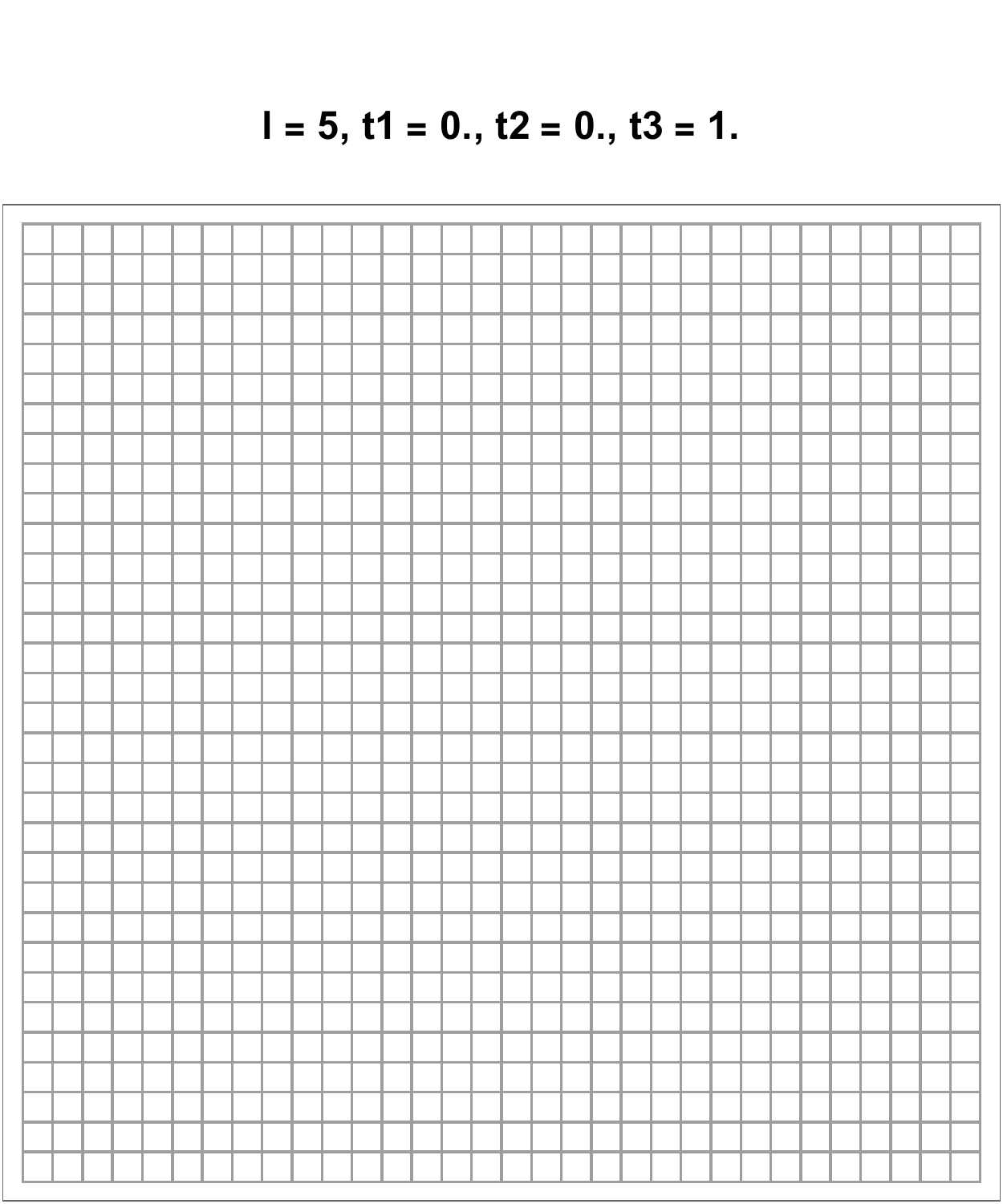}}
\end{figure}

\begin{figure}
\centering
\caption{Collection of 1064 pixels randomly distributed on a $2^5 \times 2^5 = 64  \times 64$ grid, which corresponds to a coarse-graining with five consecutive updates given $u=v=0, b=r=0.50$ and $t_1=0.14, t_2=0.11, t_3=0.75$. Each level of the coarse-graining is numbered $l= 1, 2, 3, 4, 5$ with $l=0$ for the actual collection of pixels. For each level $l$, $t_1, t_2, t_3$ are the  proportions of aggregates respectively blue, red, white. Parameters $t_1, t_2, t_3, l$ are identical to $p_n, q_n, 1-p_n-q_n, n$ used for above update equations.}
\label{a05}
\end{figure}

In the second case, I shit slightly the proportion of R to  $t_1=0.14, t_2=0.12$, which is now outside the basin of attraction. The simulation shown in Fig. (\ref{b05}) recovers a B macro-color as expected from the upper left part of Fig. (\ref{a05}). However, this final label is wrong since both $p_0$ and $q_0$ are less than $50\%$.

\begin{figure}[htbp]
\vspace{-3cm}
\centering
\makebox[\textwidth][c]{
\includegraphics[width=0.45\textwidth]{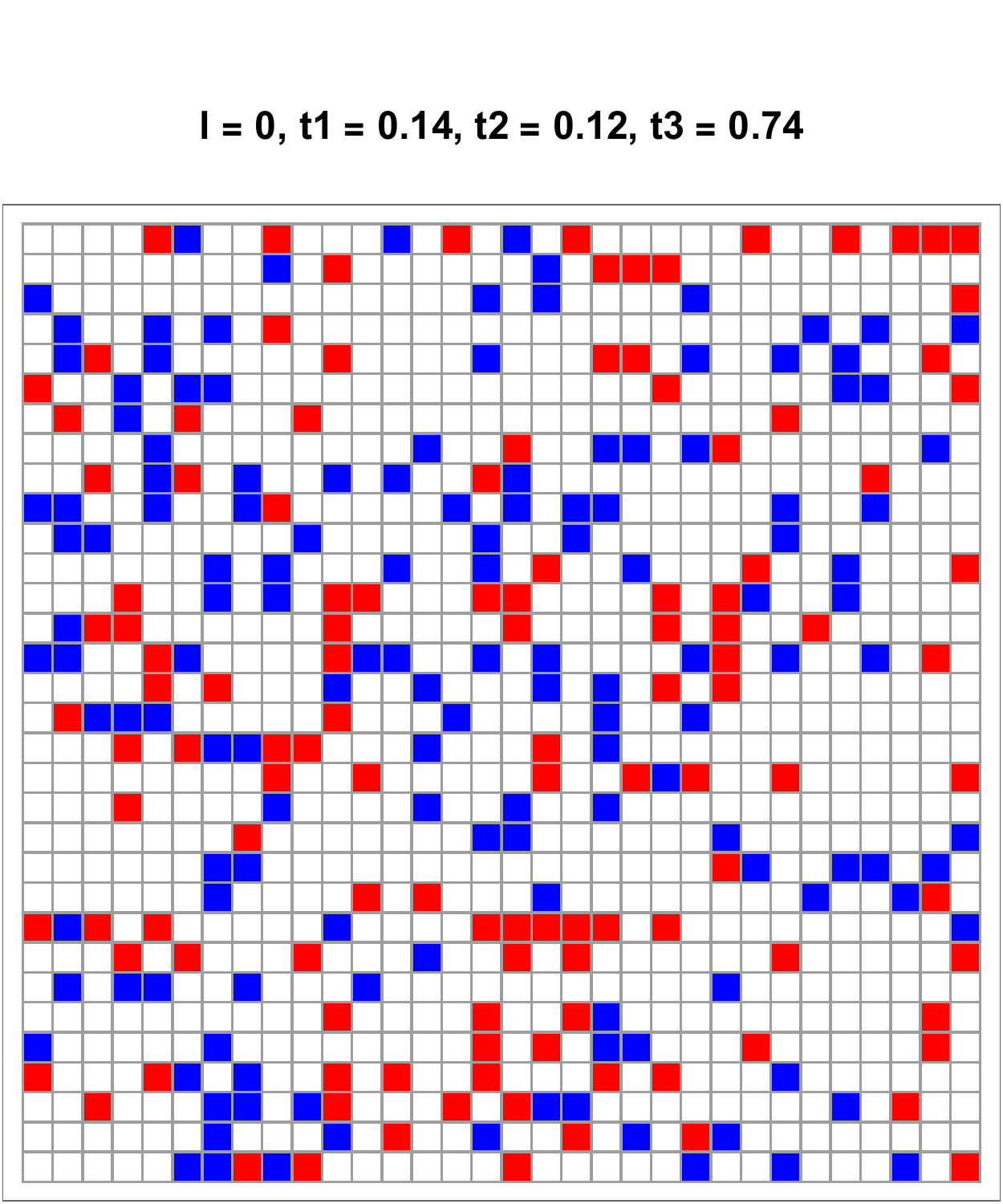}
\includegraphics[width=0.45\textwidth]{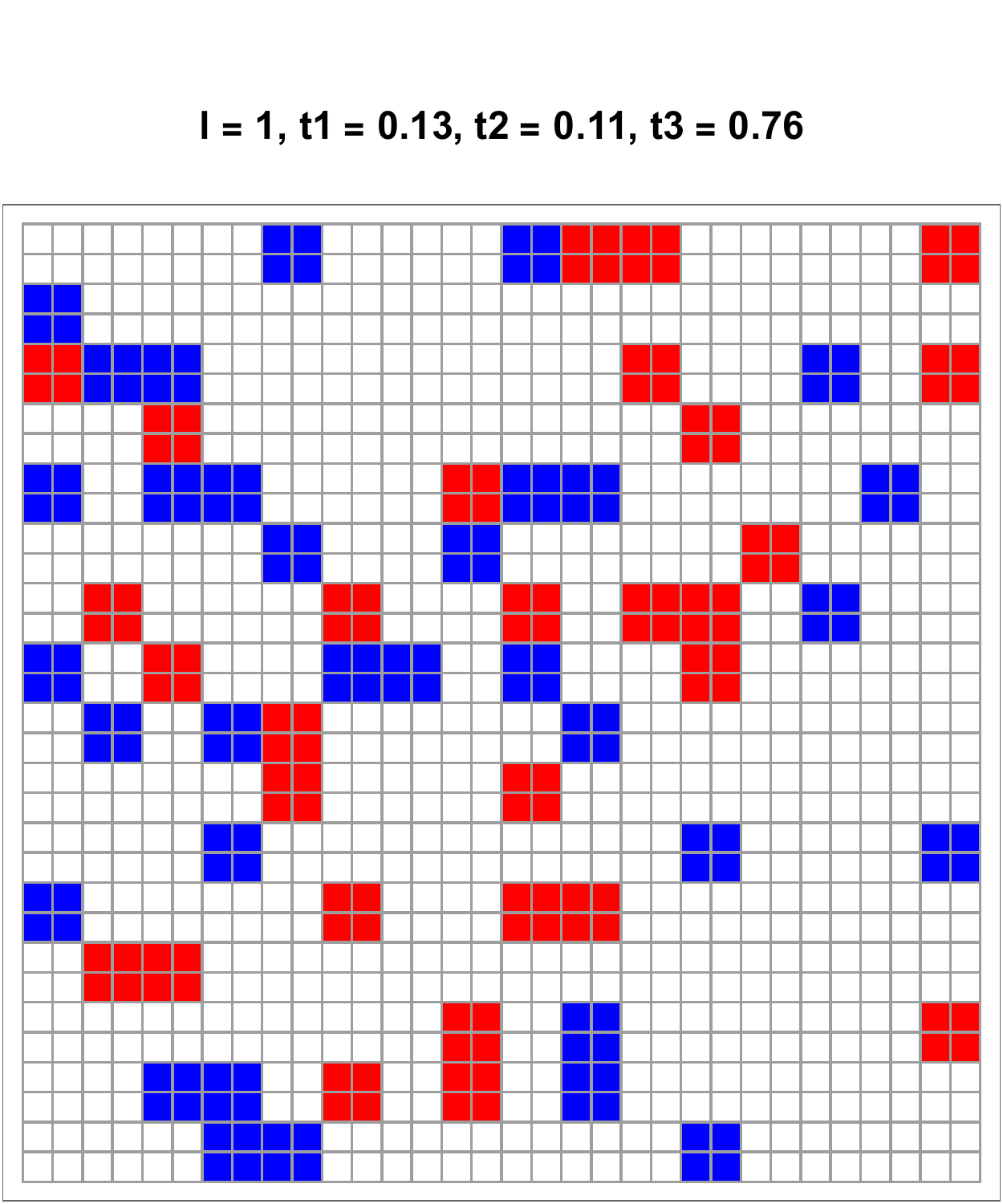}}
\vspace{0.5cm}
\makebox[\textwidth][c]{
\includegraphics[width=0.45\textwidth]{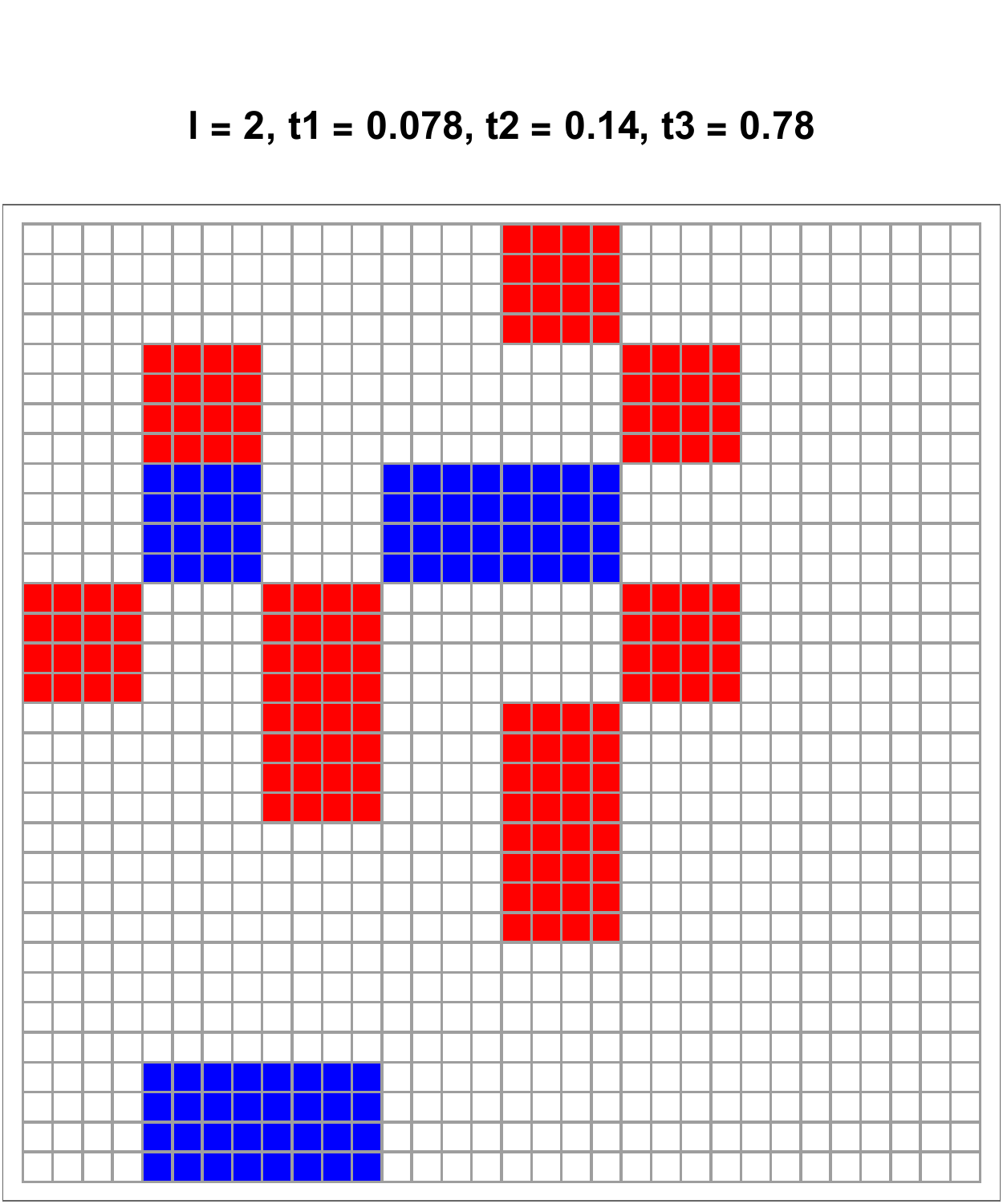}
\includegraphics[width=0.45\textwidth]{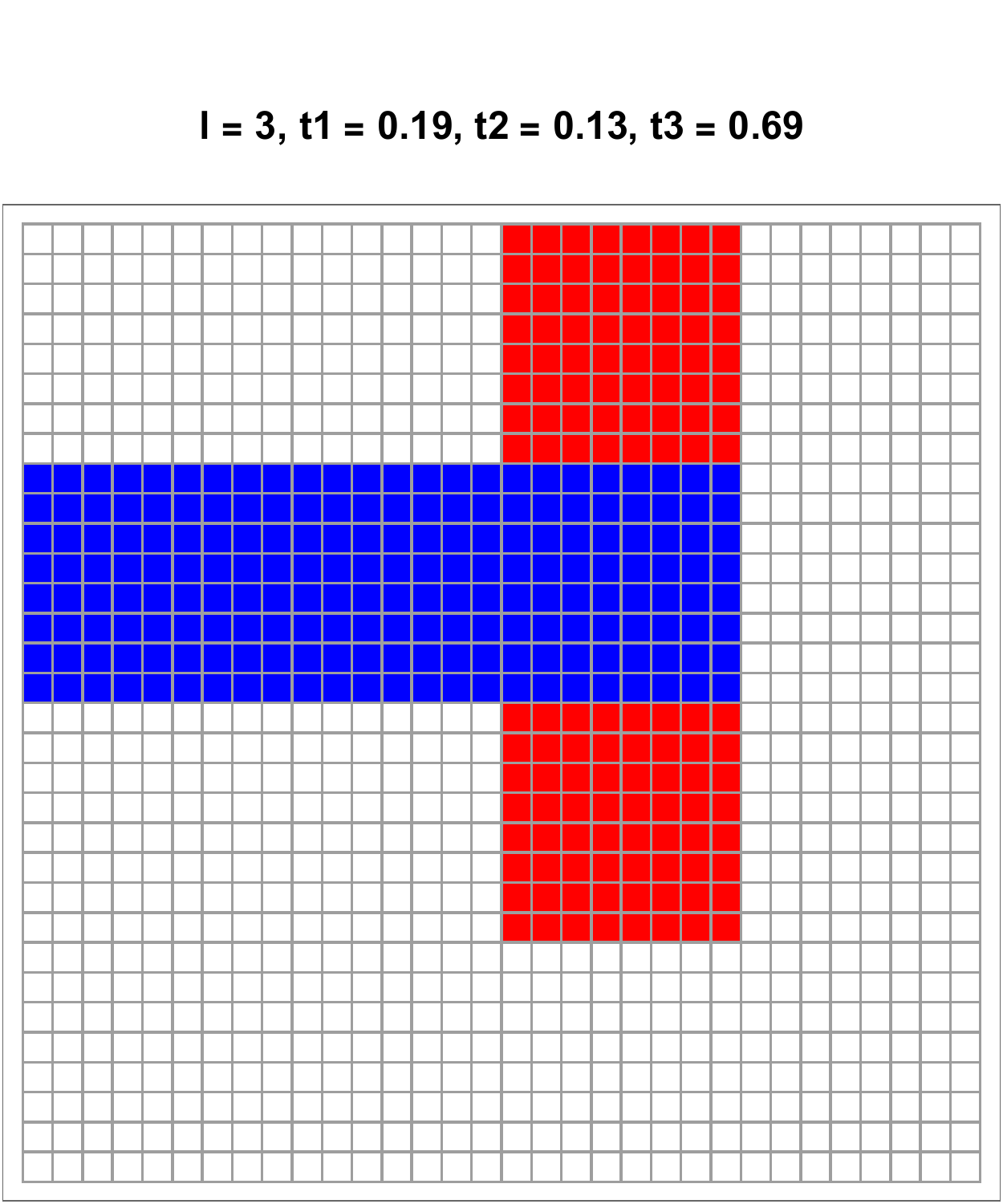}}
\vspace{0.5cm}
\makebox[\textwidth][c]{
\includegraphics[width=0.45\textwidth]{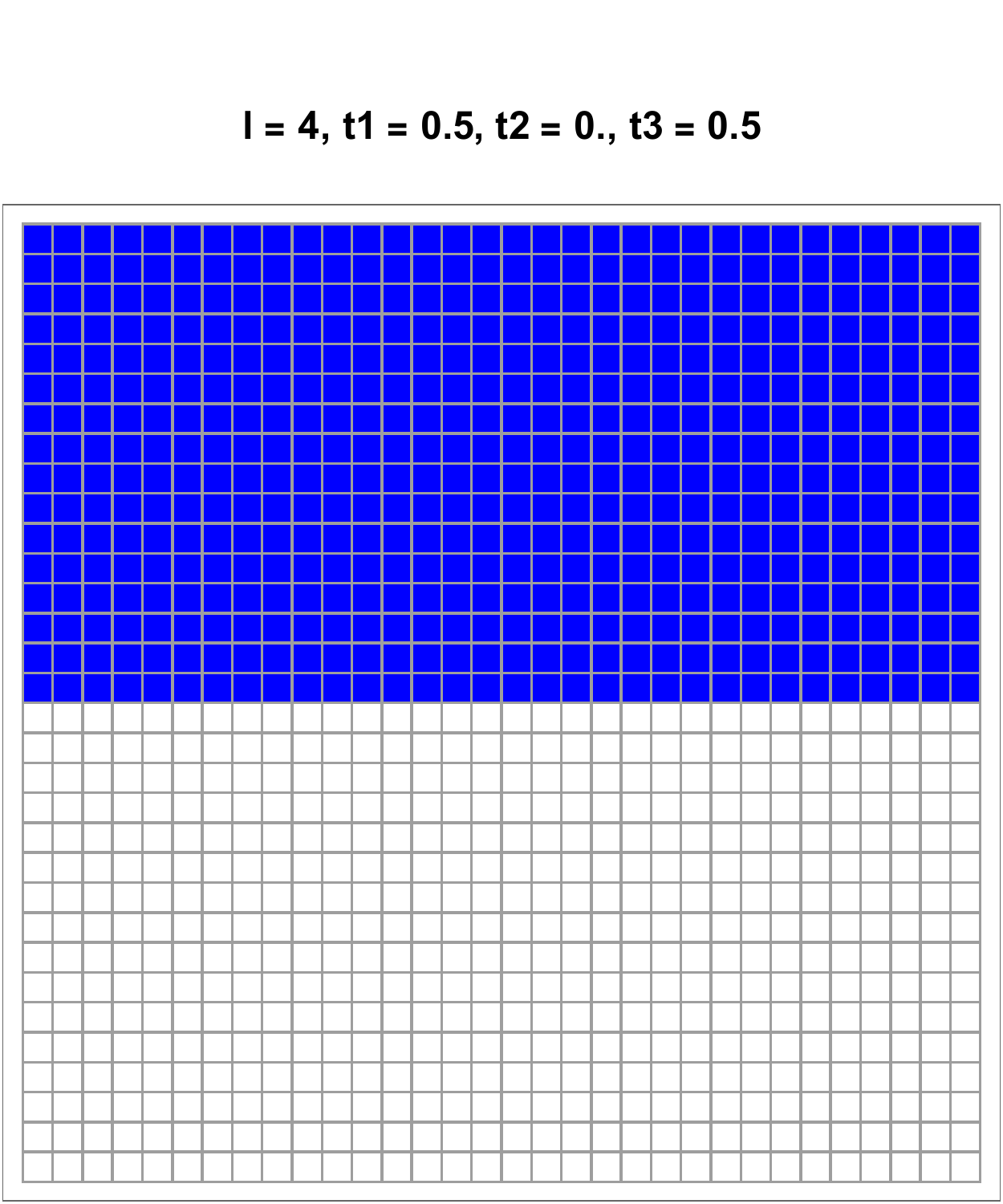}
\includegraphics[width=0.45\textwidth]{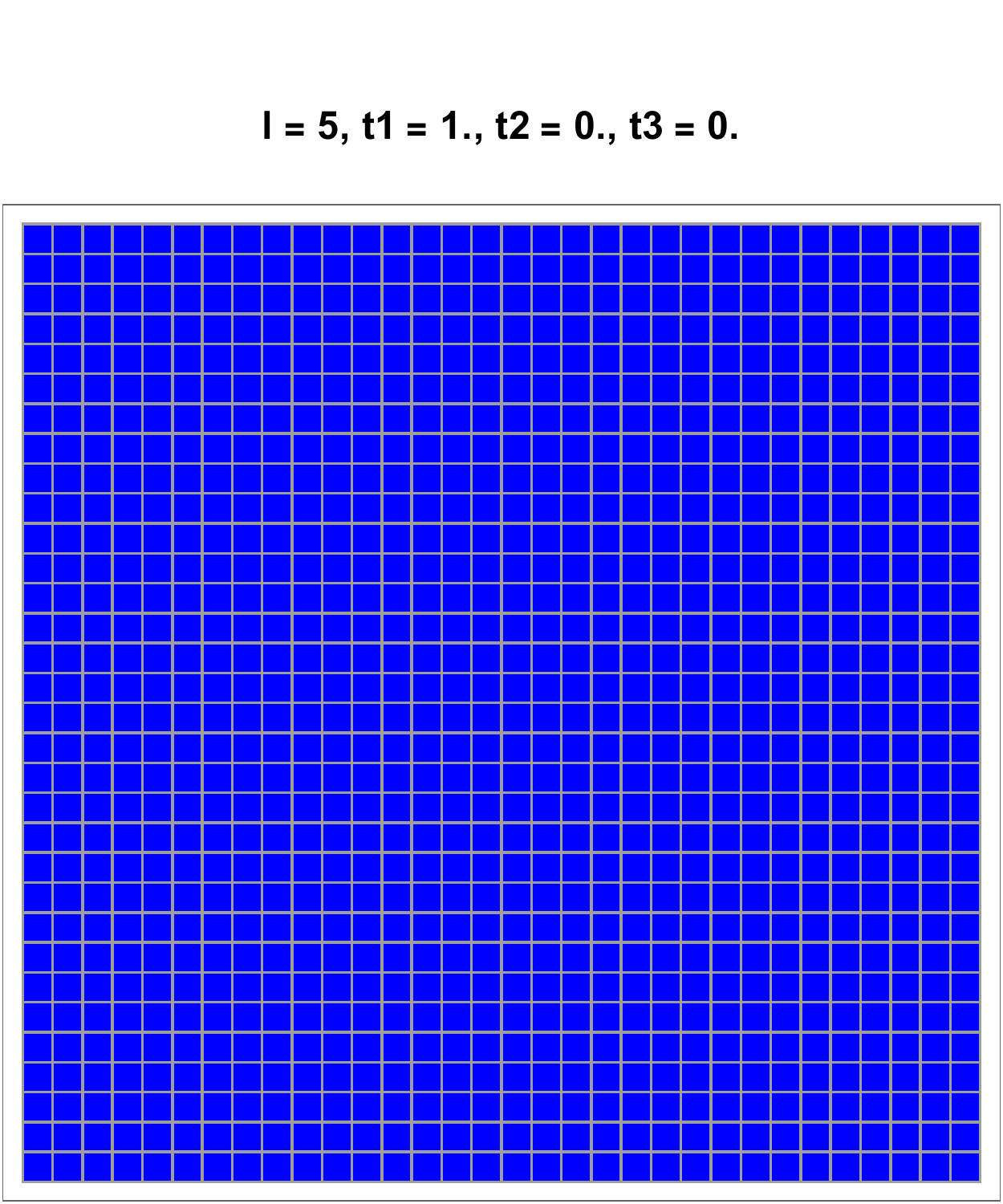}}
\end{figure}

\begin{figure}
\centering
\caption{Collection of 1064 pixels randomly distributed on a $2^5 \times 2^5 = 64  \times 64$ grid, which corresponds to a coarse-graining with five consecutive updates given $u=v=0, b=r=0.50$ and $t_1=0.14, t_2=0.12, t_3=0.74$. Each level of the coarse-graining is numbered $l= 1, 2, 3, 4, 5$ with $l=0$ for the actual collection of pixels. For each level $l$, $t_1, t_2, t_3$ are the  proportions of aggregates respectively blue, red, white. Parameters $t_1, t_2, t_3, l$ are identical to $p_n, q_n, 1-p_n-q_n, n$ used for above update equations.}
\label{b05}
\end{figure}

Last case has $u=v=1$, $b=r=0$ and $t_1=0.76, t_2=0.15$. While the exact macro-color is B, the coarse-graining yields wrongly W similarly to the update Equations as seen in the lower right part of  Fig. (\ref{f812}). The reason being that the $p$ axis has W and B attractors separated by a saddle fixed point located at 0.77.

\begin{figure}
\vspace{-3cm}
\centering
\makebox[\textwidth][c]{
\includegraphics[width=0.45\textwidth]{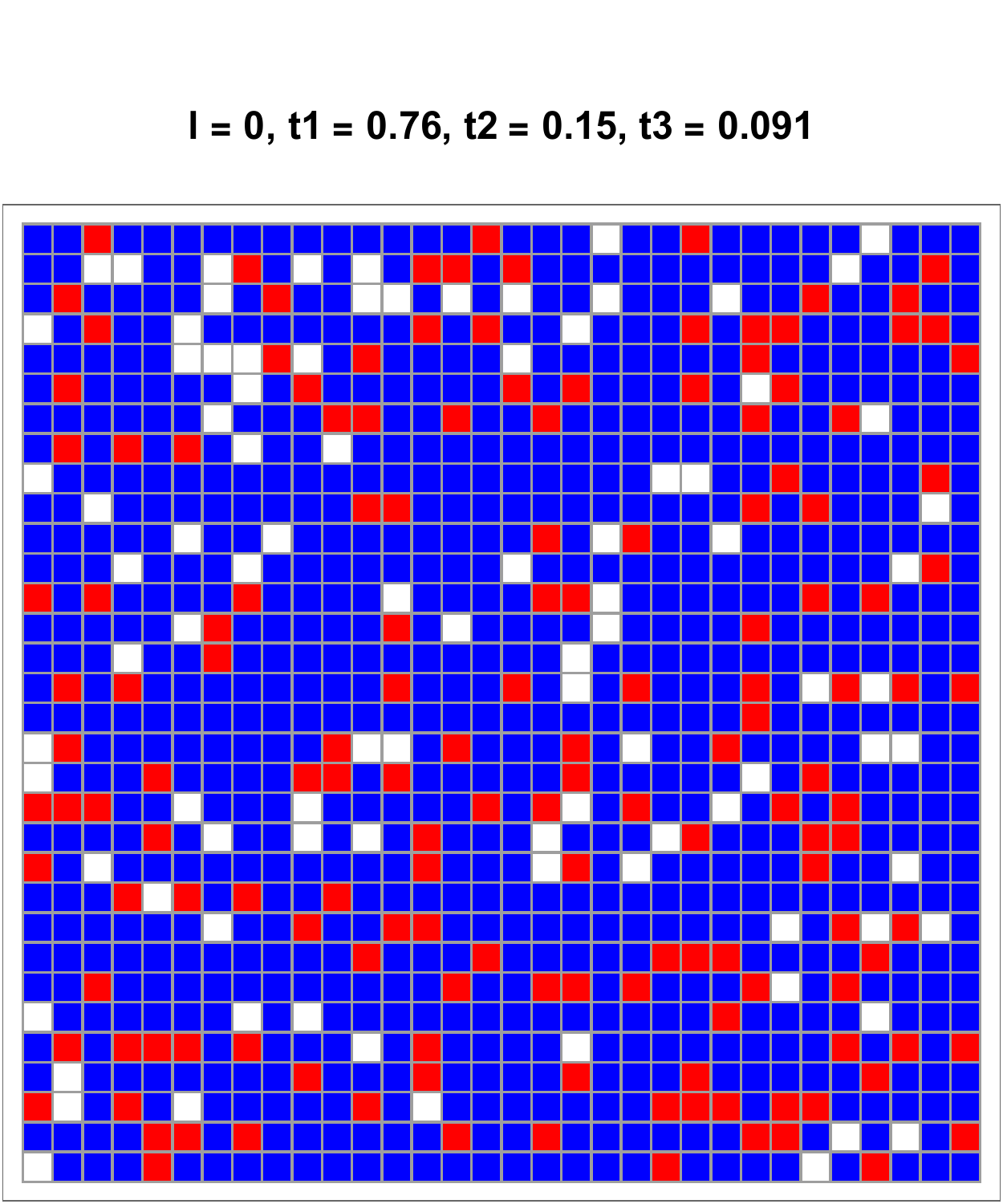}
\includegraphics[width=0.45\textwidth]{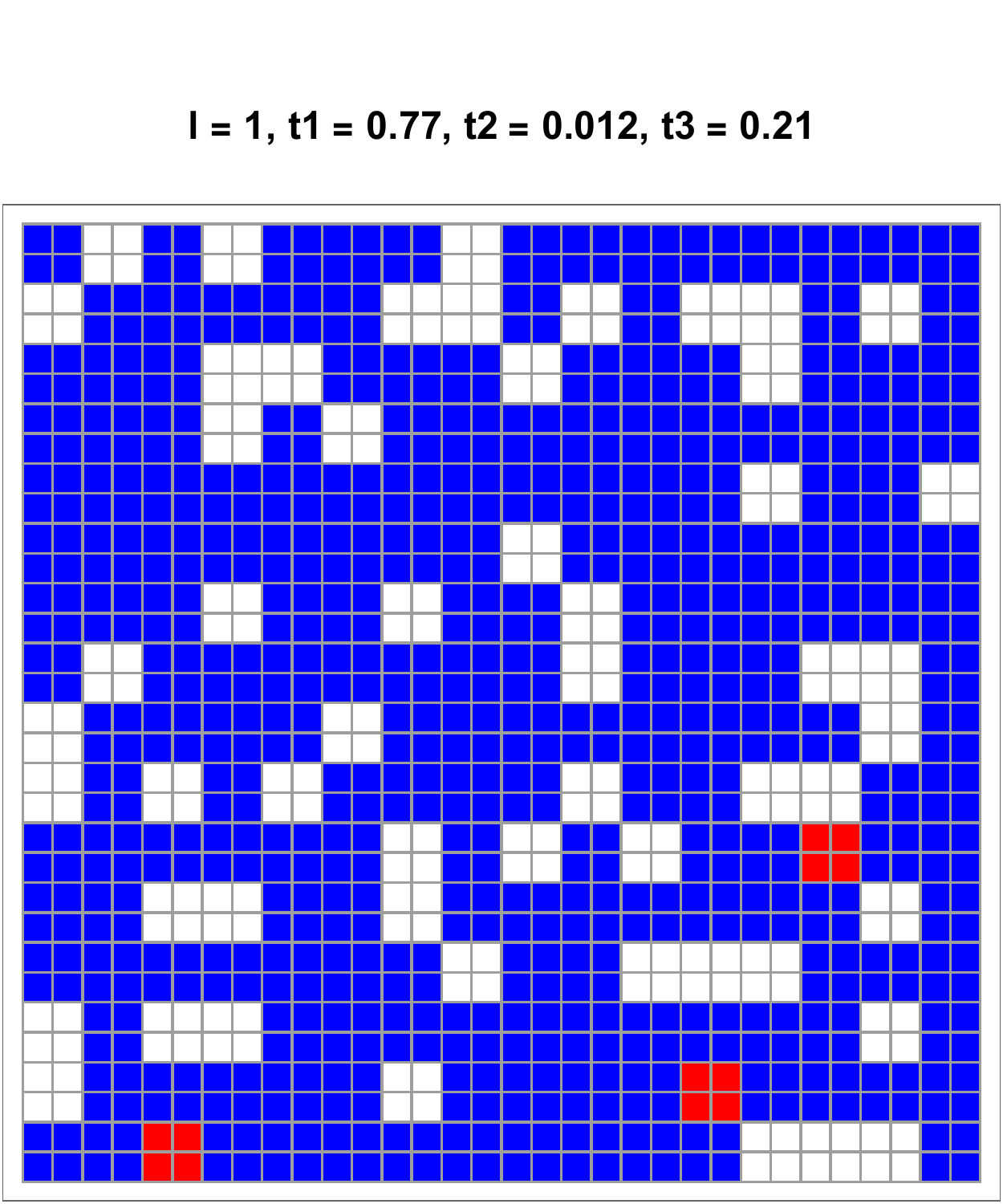}}
\vspace{0.5cm}
\makebox[\textwidth][c]{
\includegraphics[width=0.45\textwidth]{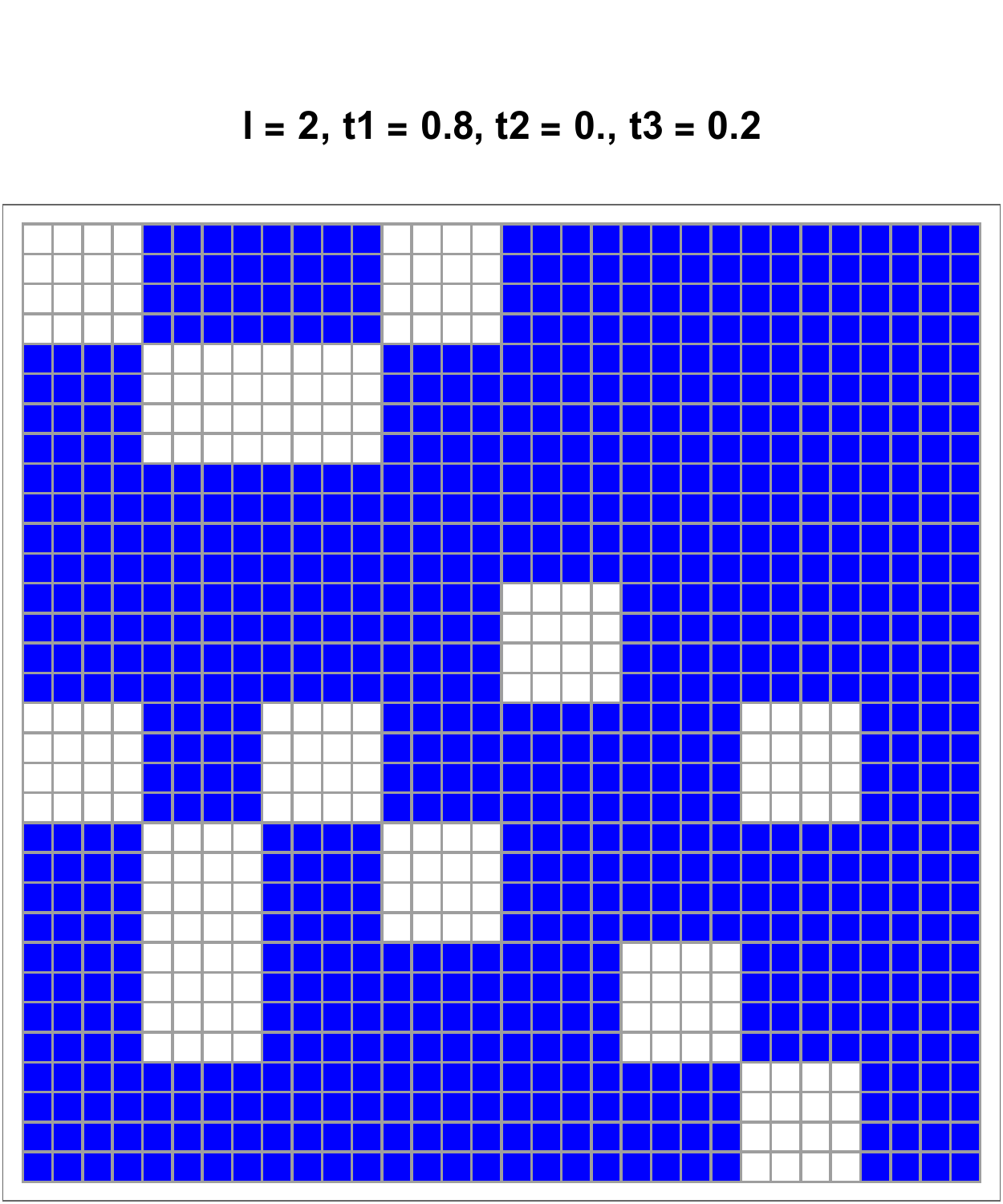}
\includegraphics[width=0.45\textwidth]{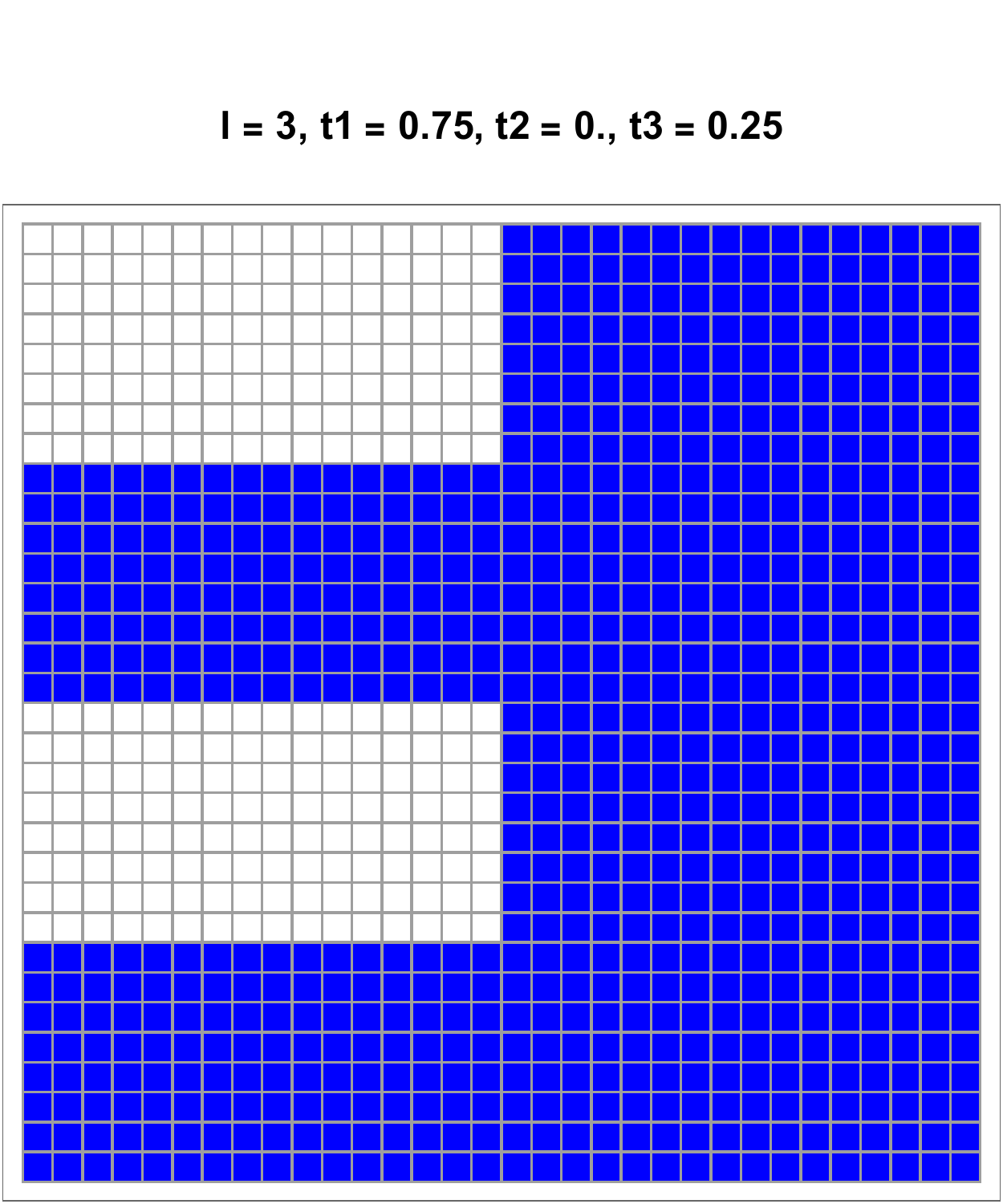}}
\vspace{0.5cm}
\makebox[\textwidth][c]{
\includegraphics[width=0.45\textwidth]{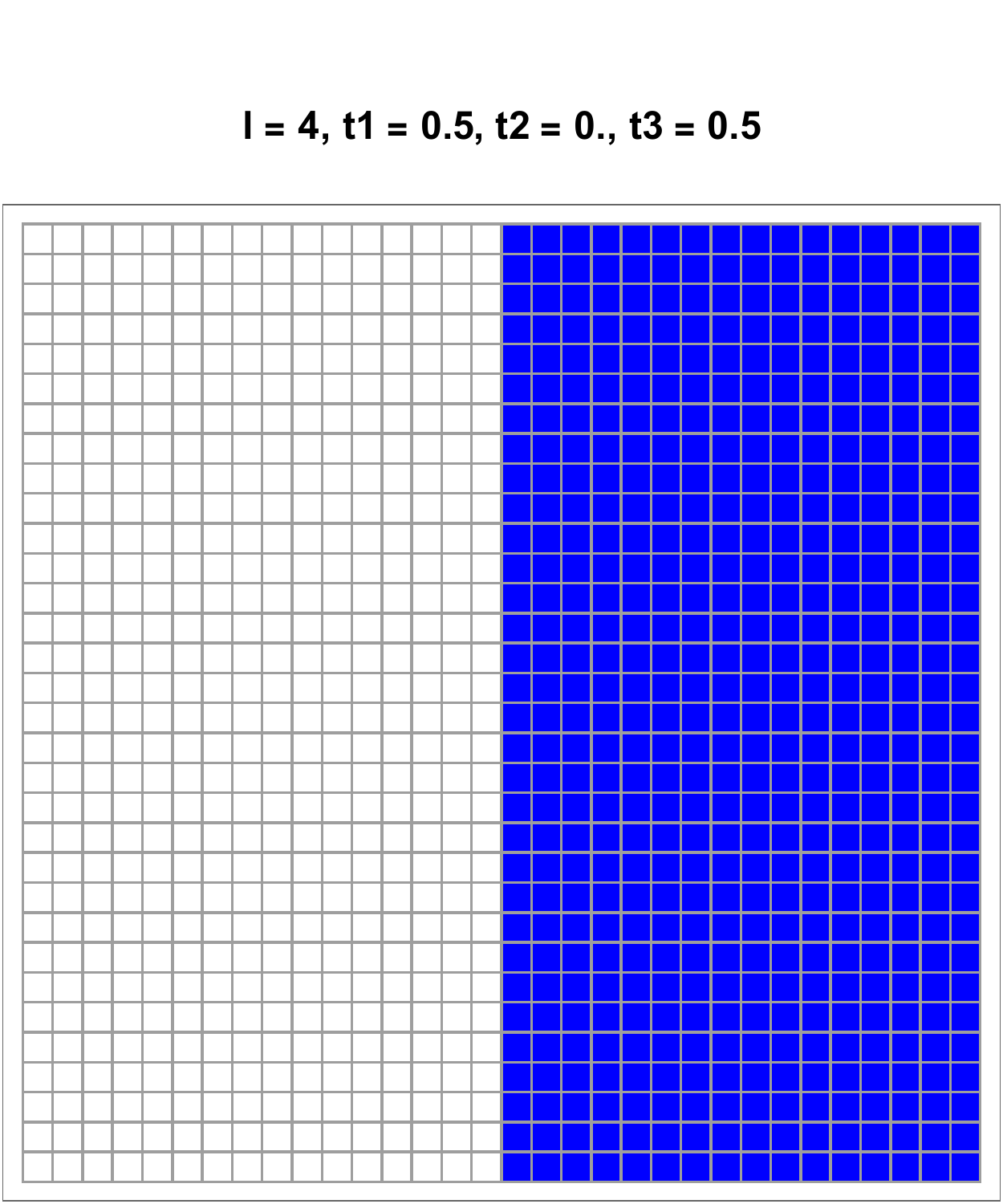}
\includegraphics[width=0.45\textwidth]{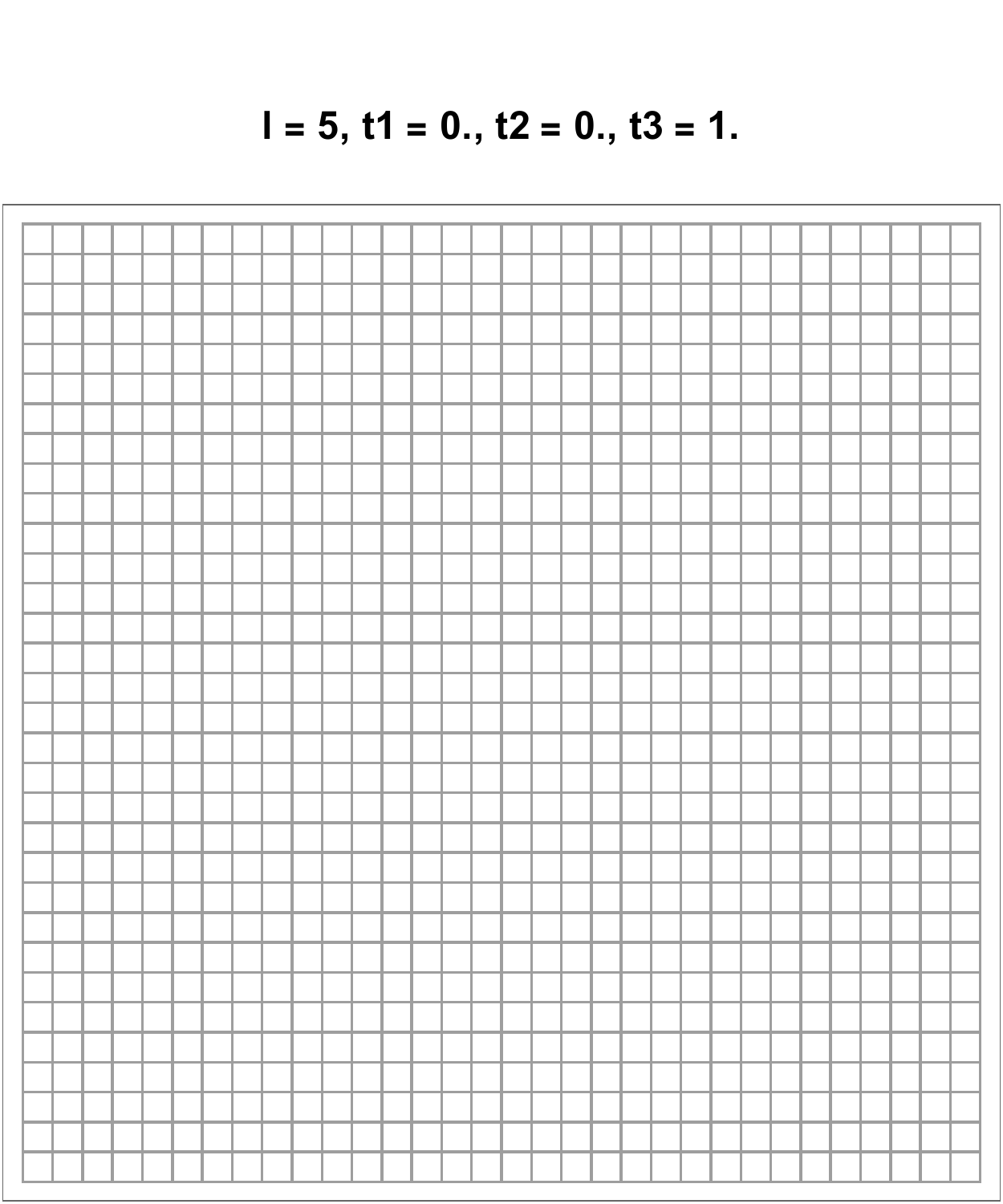}}
\end{figure}

\begin{figure}
\centering
\caption{Collection of 1064 pixels randomly distributed on a $2^5 \times 2^5 = 64  \times 64$ grid, which corresponds to a coarse-graining with five consecutive updates given $u=v=1$, $b=r=0$ and $t_1=0.76, t_2=0.15$. Each level of the coarse-graining is numbered $l= 1, 2, 3, 4, 5$ with $l=0$ for the actual collection of pixels. For each level $l$, $t_1, t_2, t_3$ are the  proportions of aggregates respectively B
blue, red, white. Parameters $t_1, t_2, t_3, l$ are identical to $p_n, q_n, 1-p_n-q_n, n$ used for above update equations.}
\label{c05}
\end{figure}

The three simulations exhibited have recovered the results yielded by the update Equations. At this stage, showing more simulations is not needed.

\section{Conclusion}

To address the subtle challenge of extracting reliable insights from large datasets, I explored the minimal yet illustrative case of a collection of colored pixels - red, blue, or white - whose overall macro-color is defined by the actual majority color among the pixels. Assuming the pixel color proportions are not directly accessible, I applied a recursive coarse-graining procedure to infer the macro-color from local configurations.

This approach frames the problem within the broader context of classification under uncertainty in big data. Each coarse-graining step acts as a local classifier operating with incomplete local information, specifically the appearance of aggregates without local majority, and the entire procedure forms a stacked ensemble of such decisions.

The analysis shows that recursive majority-vote rules frequently fail to recover the correct macro-color due to the unavoidable unclassified appearance of white aggregates. These ambiguous cases require arbitrary decisions to proceed. By systematically exploring the space of such decisions, I identified the parameter regimes in which the process yields incorrect outcomes.

The central conclusion of this work is that, regardless of the chosen local decision rules, which are inevitably required, inherent related biases propagate through the hierarchy, leading to distortions in the final classification. In this sense, the coarse-graining process itself becomes a source of systematic error, even when starting from an unbiased configuration.

In particular, the results highlight how the mere presence of unclassified white aggregates misleads the inference of global properties, underlining the fragility of hierarchical procedures in the face of local ambiguities.

This study underscores that robust insight extraction depends not only on the volume of data, but also on how local uncertainties are handled. Hierarchical aggregation schemes inherently act as symmetry-breaking mechanisms, whether through deterministic or stochastic rules, ultimately favoring one outcome over another.

Though these local  instances of symmetry breaking may appear sound, minor or inconsequential, their cumulative effect can be profound. As ambiguities propagate through the hierarchy, even small asymmetries in rule design or data distribution can give rise to significant macroscopic biases, even in complete datasets. 

Though based on a simple model, the findings offer insights into widely used data reduction techniques. In particular, they demonstrate that the impact of unintentional bias in hierarchical classification processes is unavoidable. Recognizing and addressing this practical  flaw is crucial for extracting meaningful categorical insights from complex, multi-scale data.


\end{document}